\pdfoutput=1  

\documentclass[aps,prd,superscriptaddress,floatfix,nofootinbib,preprintnumbers,eqsecnum,twocolumn]{revtex4-2}

\usepackage[normalem]{ulem}
\usepackage{amsmath,amssymb,float,graphicx,placeins,multirow}
\usepackage{enumerate,slashed,cancel,comment,color,bbm,rotating,placeins,mathtools,multirow,hhline}
\usepackage{array}
\usepackage{lipsum}

\usepackage{xkcdcols} 
\usepackage{hyperref}
\usepackage{lipsum}
\hypersetup
{
  colorlinks,
  linkcolor={blue!80!black},
  citecolor={blue!80!black},
  urlcolor={blue!80!black}
}

\newcommand{\ul}{\underline}

\newcommand{\nbox}{{\,\lower0.9pt\vbox{\hrule \hbox{\vrule height 0.2 cm
\hskip 0.2 cm \vrule height 0.2 cm}\hrule}\,}}

\newcommand{\Mpc}{\text{Mpc}}

\newcommand{\ie}{{\em i.e.}}

\newcommand{\etc}{{\it etc}.\/}

\newcommand{\kFSH}{k_{\rm FSH}}
\newcommand{\beq}{\begin{equation}}
\newcommand{\eeq}{\end{equation}}
\newcommand{\lsim}{\lower.7ex\hbox{{\il{\;\stackrel{\textstyle<}{\sim}\;}}}}
\newcommand{\gsim}{\lower.7ex\hbox{{\il{\;\stackrel{\textstyle>}{\sim}\;}}}}

\newcommand{\nn}{\nonumber}
\newcommand{\expt}[1]{\left\langle #1 \right\rangle}
\newcommand{\il}[1]{\mbox{$#1$}}
\newcommand{\abs}[1]{\left| #1 \right|}

\newcommand{\vj}[1]{\langle v \rangle_{\hspace*{-0.2mm}#1}}
\newcommand{\pj}[1]{\langle p \rangle_{\hspace*{-0.2mm}#1}}
\newcommand{\kmin}{k_{\rm min}}
\newcommand{\kmax}{k_{\rm max}}
\newcommand{\mWDM}{{m_{\rm WDM}}}
\newcommand{\TWDM}{T_{\rm WDM}}
\newcommand{\mdA}{m_{\delta A}}
\newcommand{\khalf}{{k_{\hspace{-0.07mm}1\hspace{-0.17mm}/\hspace{-0.08mm}2}}}
\newcommand{\mhalf}{{m_{\hspace{-0.07mm}1\hspace{-0.17mm}/\hspace{-0.08mm}2}}}

\newcommand{\dA}{\delta A}
\newcommand{\dAWDM}{\delta A_{\rm WDM}}
\newcommand{\Td}{T_{\hspace{-0.2mm}d}}
\newcommand{\ad}{a_{\hspace{-0.2mm}d}}
\newcommand{\gint}{g_{\rm int}}

\def\calN{{\cal N}}
\def\calO{{\cal O}}
\def\tnow{t_{\mathrm{now}}}

\def\Lya{{Lyman-$\alpha$}}

\def\tMRE{{t_{\rm MRE}}}

\graphicspath{{./figs}}

\begin{document}
\title{Evaluating Lyman-$\alpha$ Constraints for General Dark-Matter Velocity Distributions:\\
Multiple Scales and Cautionary Tales}

\def\andname{\hspace*{-0.5em}} 
\author{Keith R. Dienes}
\email[Email address: ]{dienes@arizona.edu}
\affiliation{Department of Physics, University of Arizona, Tucson, AZ 85721 USA}
\affiliation{Department of Physics, University of Maryland, College Park, MD 20742 USA}
\author{Fei Huang}
\email[Email address: ]{huangf4@uci.edu}
\affiliation{Department of Physics and Astronomy, University of California, Irvine, CA  92697 USA}
\affiliation{CAS Key Laboratory of Theoretical Physics, Institute of Theoretical Physics, Chinese Academy of Sciences, Beijing 100190 China}
\author{Jeff Kost}
\email[Email address: ]{j.d.kost@sussex.ac.uk}
\affiliation{\mbox{Department of Physics \& Astronomy, University of Sussex, Brighton BN1 9QH, United Kingdom}}
\author{Brooks Thomas}
\email[Email address: ]{thomasbd@lafayette.edu}
\affiliation{Department of Physics, Lafayette College, Easton, PA 18042 USA}
\author{Hai-Bo Yu}
\email[Email address: ]{haiboyu@ucr.edu}
\affiliation{\mbox{Department of Physics and Astronomy, University of California, Riverside, CA  92521 USA}}


\begin{abstract}
One of the most important observational constraints on possible 
models of dark-matter physics exploits the Lyman-$\alpha$ absorption 
spectrum associated with photons traversing the intergalactic medium.
Because this data allows us to probe the linear matter power spectrum with 
great accuracy down to relatively small distance scales, finding ways of 
accurately evaluating such Lyman-$\alpha$ constraints across large classes
of candidate models of dark-matter physics is of paramount importance.  While 
such Lyman-$\alpha$ constraints have been evaluated for dark-matter models 
that give rise to relatively simple dark-matter velocity distributions, 
more complex models --- particularly those whose dark-matter velocity 
distributions stretch across multiple scales --- have recently been 
receiving increasing attention.  In this paper, we undertake a study 
of the Lyman-$\alpha$ constraints associated with general dark-matter 
velocity distributions.  Although such Lyman-$\alpha$ constraints are 
difficult to evaluate in principle, in practice there currently exist 
two classes of methods in the literature through which such constraints 
can be recast into forms which are easier to evaluate and which therefore 
allow a more rapid determination of whether a given 
dark-matter model is ruled in or out.
Accordingly, we utilize both of these recasts in order to determine the 
Lyman-$\alpha$ bounds on different classes of dark-matter velocity 
distributions.  We also develop a general method by which the results 
of these different recasts can be compared.  For relatively simple 
dark-matter velocity distributions, we demonstrate that these two 
classes of recasts tend to align and give similar results.  However, we 
find that the situation is far more complex for distributions involving 
multiple velocity scales: while these two classes of recasts continue 
to yield similar results within certain regions of parameter space, they
nevertheless yield dramatically different results within precisely those 
regions of parameter space which are likely to be phenomenologically 
relevant. This, then, serves as a cautionary tale regarding the use of such 
recasts for complex dark-matter velocity distributions.    
\end{abstract}

\maketitle


\section{Introduction\label{sec:Introduction}}


On large scales, the $\Lambda$CDM cosmology --- a cosmology which is
built on a relatively simple set of assumptions and whose components include only 
ordinary (visible) matter, cold dark matter (CDM), and a cosmological constant 
$\Lambda$ --- has provided a remarkably successful model of the 
universe~\cite{Planck:2018vyg}.  However, on smaller scales corresponding to 
distances \il{\lesssim 1\, \Mpc}, evidence of this success is not as solid, with 
inconsistencies between numerical simulations and
observations~\cite{DelPopolo:2016emo,Tulin:2017ara,Bullock:2017xww,Perivolaropoulos:2021jda} appearing in several areas.

Several of these inconsistencies can potentially be addressed by positing that the 
dark matter is not strictly cold.  Indeed, there exist a variety of mechanisms 
through which a population of dark-matter particles with a non-negligible 
primordial velocity distribution can be produced in the early universe.  However, 
different mechanisms can give rise to different velocity distributions with 
dramatically different shapes.  

\begin{figure}[t]
    \begin{center}
    \includegraphics[width=0.49\textwidth, keepaspectratio]{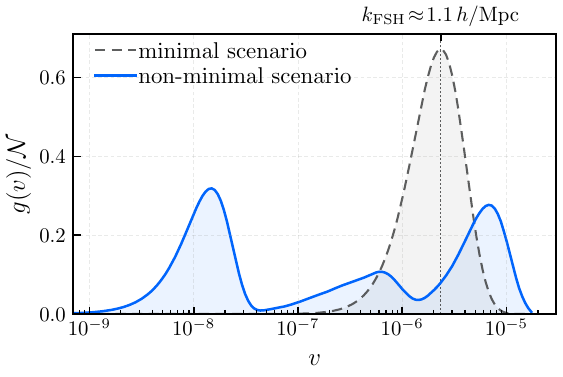}
        \caption{
            Examples of possible dark-matter velocity distributions $g(v)/\mathcal{N}$ 
            which lead to departures from the small-scale structure 
            predicted by CDM.~  The dashed gray curve represents the 
            distribution which arises from the thermal freeze-out of 
            a dark-matter particle with a mass of 
            \il{\mathcal{O}({\rm keV})}.  By contrast, the solid 
            blue curve represents a highly non-thermal velocity 
            distribution generated by decay cascades within a non-minimal 
            dark sector~\cite{Dienes:2020bmn}.  Both of the 
            distributions shown have the same present-day average velocity
            and would na\"{i}vely be characterized by the same 
            free-streaming horizon \il{k_{\rm FSH}\approx 1.1\, h/\Mpc}.
            }
            \label{fig:introdist}
    \end{center}
\end{figure}

In order to illustrate the range of
possibilities, in Fig.~\ref{fig:introdist} we provide two examples of 
possible dark-matter velocity distributions which arise in different 
dark-matter scenarios.
The first of these distributions (dashed gray curve) arises from a warm dark 
matter (WDM) scenario wherein a dark-matter particle with a mass 
of \il{\mathcal{O}({\rm keV})} is produced via thermal freeze-out while still 
relativistic.  It is a simple, unimodal distribution which is sharply peaked 
around a particular velocity.  The second distribution (solid blue curve), by 
contrast, arises within the context of a dark-matter
model~\cite{Dienes:2020bmn} in which the dark matter is produced as 
the end-product of a long 
sequence of possible decays that take place within a tower of heavier, 
unstable dark states.  The non-trivial shape of the resulting dark-matter 
velocity distribution then reflects the contributions of many possible decay 
chains within such a scenario.  Indeed, the resulting 
dark-matter velocity distribution can be highly complex and even multi-modal.
We emphasize that both of the
distributions shown in Fig.~\ref{fig:introdist} have the same average velocity
and would na\"{i}vely be characterized by the same free-streaming horizon 
\il{k_{\rm FSH}\approx 1.1\, h/\Mpc}, where $h$ is the dimensionless 
Hubble constant.  However, they also clearly differ significantly and 
indeed give rise to different predictions for small-scale structure --- even 
for the same background cosmology.

One way in which the detailed structure of the primordial dark-matter velocity 
distribution leaves its imprint on small-scale structure is through 
free-streaming effects, which impact the shape of the linear matter power 
spectrum.  Indeed, a great deal of information about the velocity distribution 
can be inferred from the shape of the matter power spectrum directly.
That said, deriving rigorous constraints on the dark-matter velocity 
distribution in this way presents some challenges.  One such challenge 
stems from the fact that the \Lya\ forest currently provides the most 
stringent constraints on the shape of the matter power spectrum across a 
broad range of comoving wavenumbers $k$.  In order to translate 
these \Lya\ constraints 
into rigorous bounds on any particular dark-matter model, one would 
typically need to perform detailed numerical hydrodynamic simulations specifically 
tailored to that model and its characteristic dark-matter velocity distribution.  
Since simulations of this sort are computationally intensive, 
it is impractical to employ them when surveying broad classes 
of dark-matter models.  

For this reason, methods for evaluating 
these constraints have been developed which effectively involve 
{\it recasting}\/ the \Lya\ data onto a baseline WDM model for which 
these sorts of hydrodynamic simulations have already been carried out.  
Different methods exist in the literature for constructing such recasts, 
but these methods typically fall into two general classes.  The first 
class comprises methods which are sensitive to the properties of the 
power spectrum only at relatively low values of 
$k$~\cite{Konig:2016dzg,Heeck:2017xbu,Bae:2017dpt,Boulebnane:2017fxw,Calibbi:2018fqf,Garny:2018ali,Kamada:2019kpe,Baldes:2020nuv,Dvorkin:2020xga,Ballesteros:2020adh,DEramo:2020gpr,Baumholzer:2020hvx}.
By contrast, the second class comprises methods which are sensitive to the 
properties of $T(k)$ over a broader range of 
$k$~\cite{Schneider:2016uqi,Murgia:2017lwo,Irsic:2017ixq,Kobayashi:2017jcf,Huo:2017vef,Murgia:2017cvj,Chacko:2018vss,Murgia:2018now,Bae:2019sby,Huo:2019bjf,Buen-Abad:2021mvc}.
We shall refer to recasts which fall into these two classes as 
``half-mode'' and $\dA$ recasts, respectively.
The methods involved in formulating recasts within each class will be 
reviewed in Sect.~\ref{sec:methods}.  

In this paper, we shall utilize these recasts in order to undertake a 
general study of \Lya\ constraints on dark-matter models which 
transcend the traditional CDM hypothesis.  More specifically, 
we shall utilize these recasts in order to evaluate the \Lya\ 
constraints for a variety of dark-matter velocity distributions.
Given that both half-mode and $\dA$ recasts are constructed in reference 
to a baseline WDM velocity distribution, we shall find it useful to 
classify the different possible dark-matter velocity distributions that 
we shall examine in this paper into two overall categories:
\begin{itemize}

\item \ul{Class~I}: velocity distributions whose shapes are well approximated 
  by the characteristic profile which arises in WDM scenarios.  Such velocity 
  distributions have a single peak, as exemplified by the unimodal distribution 
  depicted in Fig.~\ref{fig:introdist}, and rise and fall smoothly 
  from that peak on either side. 

\item \ul{Class~II}: more complicated velocity distributions whose shapes depart 
  significantly from the WDM profile.  Such distributions exhibit a variety of 
  different internal features, and include multi-modal distributions of the sort 
  illustrated in Fig.~\ref{fig:introdist}. 
 
\end{itemize}

We shall begin our analysis by examining the \Lya\ constraints on dark-matter 
scenarios whose characteristic velocity distributions unambiguously fall within 
Class~I.  By definition, these include the baseline WDM scenario itself.  We 
shall then systematically study departures from this baseline scenario and 
investigate how the \Lya\ constraints vary as the shape of the dark-matter 
velocity distribution is modified.  Finally, fully breaking from Class~I, we 
examine the constraints on dark-matter scenarios which give rise to multi-modal 
velocity distributions --- distributions which decidedly fall within Class~II.

Because we shall employ both the half-mode and $\delta A$ recasts in 
our analysis, the results of our paper will also enable us to {\it compare}\/ 
these two classes of recasts.  For Class-I velocity distributions, we 
find that these recasts tend to align and give similar results.  
However, we find that the situation is far more complex for Class~II
velocity distributions.  While the two classes of recasts continue to 
yield similar results within certain regions of parameter space, they 
can nevertheless yield dramatically different results within precisely 
those regions of parameter space which are likely to be phenomenologically 
relevant.  Indeed, these differences can be quite significant.  
Our results thus provide a cautionary tale regarding the use of such recasts 
when more general velocity distributions are considered.    

This paper is organized as follows.  In Sect.~\ref{sec:methods}, we begin 
by reviewing the calculations necessary in order to 
evaluate the \Lya\ constraints on dark-matter models.  We also outline
the two classes of recasts that we consider in this paper.
In Sect.~\ref{sec:recasting}, we then develop a general method 
through which these two recasts can be compared on an equal footing.
In Sect.~\ref{sec:minimal}, we evaluate the \Lya\ constraints 
obtained from both of these recasts for a simple baseline dark-matter model
as well as for certain extensions of this model.
In Sect.~\ref{sec:nonminimal}, we then turn our attention to non-minimal
dark-matter scenarios in which the dark-matter velocity distribution 
is multi-modal and compare the results obtained from our two recasts within
such scenarios. In Sect.~\ref{sec:abg}, we apply a parametric function for multi-modal distributions. Finally, in Sect.~\ref{sec:Conclusions}, we discuss 
the potential implications of our results and identify areas 
meriting further exploration.

\section{Evaluating \Lya\ Constraints on Dark Matter \label{sec:methods}}


Given a proposed model of dark-matter physics and a given background 
cosmology, one can in principle calculate the 
spatial distribution of dark matter within and between galaxies.  The spatial 
distribution of dark matter affects the spatial distribution of visible 
matter --- including the frequency and density of clouds of neutral 
hydrogen --- which can be probed by astronomical observations.  In 
particular, as light emanating from distant quasars and other bright 
objects passes through the intergalactic medium, it
interacts with the neutral hydrogen and can be observed today.
By analyzing the \Lya\ absorption spectrum of this radiation,
one can infer properties of the hydrogen distribution.
In this way, measurements of the observed \Lya\  absorption spectrum 
within such light can indirectly be used to constrain models of 
dark-matter physics.

As we shall see, the central object in evaluating such \Lya\ constraints 
is the so-called ``transfer function'' $T(k)$.  For any model of 
dark-matter physics with a predicted linear matter power spectrum $P(k)$, 
or for any set of cosmological observations that yield an observed 
linear matter power spectrum $P(k)$, the corresponding transfer 
function $T(k)$ is defined as
\beq
   T^2(k) ~\equiv~ \frac{P(k)}{P_{\rm CDM}(k)} ~
\eeq
where $P_{\rm CDM}(k)$ is the matter power spectrum that would have arisen 
in a model consisting of purely cold dark matter and where $k$ once again 
denotes a comoving wavenumber.  The transfer function thus tracks how 
cosmological structure, either in theoretical predictions or in data 
gathered from observations, deviates from that of CDM.

In general, in order to evaluate \Lya\ constraints on a given 
dark-matter model, one is faced with two tasks.  The first is to 
determine the transfer function $T(k)$ predicted by the theoretical
dark-matter model; the second is to translate observational data in 
such a way that it may be compared directly to these predictions.
We shall now provide a brief outline of how these tasks are typically 
performed.

\subsection{Determining Theoretical Predictions for $T(k)$\label{subsec:Pk}}

In general, the form of $T(k)$ which arises from any
particle-physics model of dark matter depends on a variety of factors.
One of the most important is the dark-matter phase-space distribution $g(p,t)$, 
which in an approximately homogeneous and isotropic universe can be 
expressed in terms of the three-momentum magnitude 
$p\equiv \abs{\vec p\hspace{0.2mm}}$ and the time $t$ alone.  
We shall find it convenient to define $g(p,t)$ as a 
distribution in $(\log p)$-space according to the relation
\beq
    N(t) ~=~ {\gint\over 2\pi^2}  \int_{-\infty}^{\infty} 
    \!\!d\log p~ g(p,t) \ ,
    \label{eq:Ndef}
\eeq
where $N(t)$ is the comoving number density of the particle species 
which constitutes the dark matter and where $\gint$ is the number of 
internal degrees of freedom for that species.  With this definition for 
$g(p,t)$, cosmological redshifting does not alter the {\it shape}\/ of 
$g(p,t)$, but instead merely shifts the entire distribution uniformly to 
lower values of $\log p$. 

For many dark-matter models, $g(p,t)$ effectively ceases evolving --- except 
in the trivial way discussed above, as a consequence of cosmological 
redshifting --- well before the time $\tMRE$ of matter-radiation 
equality.  For such models, one may unambiguously characterize the primordial 
dark-matter velocity distribution in terms of the distribution 
\il{g(p) \equiv g(p,\tnow)}, where the extrapolation of $g(p,t)$ to the 
present time $\tnow$ is performed accounting for cosmological 
redshifting alone.  Indeed, this $g(p)$ distribution tells us whether 
the dark matter is cold or hot, 
thermal or non-thermal, and so forth.  Moreover, the functional form for 
$g(p)$ contains all of the information we require in order to calculate the 
corresponding transfer function directly by evolving the spectrum of 
primordial cosmological perturbations forward in time.  In this paper, 
we shall carry out calculations of this sort using the \texttt{CLASS} software 
package~\cite{Lesgourgues:2011re,Blas:2011rf,Lesgourgues:2011rg,Lesgourgues:2011rh}.

Since $g(p)$ contains all of the information necessary to produce predictions 
for $T(k)$, we can survey large classes of dark-matter models simply by 
examining different possible profiles for $g(p)$ without worrying about the 
underlying physics from which these distributions might ultimately have been 
produced.  Indeed, while many details of the underlying physics leave unique 
imprints on the $g(p)$ distributions, many features do not.   
This issue is studied in some detail in Ref.~\cite{Dienes:2020bmn}.

Our primary interest in this paper is to investigate how dark-matter models
with different $g(p)$ distributions are constrained by \Lya\ data.  As such, 
we are primarily interested in $g(p)$ distributions for which free-streaming 
effects have an impact on structure on scales that can be meaningfully 
probed using such data.  For a Class-I distribution, the effect of 
free-streaming on $P(k)$ is comparatively straightforward to assess.  
The primary figure of merit in this regard is the present-day value 
\il{\expt{v} = \expt{v(\tnow)}} of the time-dependent average velocity 
\begin{equation}
    \expt{v(t)}\! ~=~
      \frac{2\pi^2}{g_{\rm int}}\int_{-\infty}^{\infty} 
      d\log p\, \frac{p}{\sqrt{p^2+m^2a^2(t)}}\,g(p)~,~
\end{equation}
where $a(t)$ is the scale factor, defined such that \il{a(\tnow)=1}.  In 
particular, for such distributions, one may sensibly define the 
free-streaming horizon
\begin{align}
    d_{\rm FSH} ~&\equiv~ \int \frac{dt}{a(t)}\expt{v(t)} \nn\\
    ~&\approx~ \frac{\expt{v}}{(a^2H)_{\rm MRE}}
    \left[2 + \log\left(\!\frac{2a_{\rm MRE}}{\expt{v}}\!\right)\right] \ ,
    \label{eq:dFSH}
\end{align}
where the subscript ``MRE'' represents the value of the corresponding quantity 
at $\tMRE$.  Roughly speaking, the corresponding wavenumber 
\il{\kFSH\sim 1/d_{\rm FSH}} represents the value of $k$ above which $P(k)$ is
suppressed by free-streaming effects for a given $\expt{v}$.  Given that   
\il{(a^2H)_{\rm MRE} \approx 8\times 10^{-6}\,h/\Mpc} and 
\il{a_{\rm MRE}\approx 2.84\times 10^{-4}}, we find that 
the range of velocities which correspond to the range of wavenumbers
\il{1\,h/\Mpc \lesssim k \lesssim 50\,h/\Mpc} constrained by \Lya\ data is
roughly \il{10^{-8} \lesssim \expt{v} \lesssim 10^{-6}}.  For 
Class-I dark-matter velocity distributions, then, assessing whether 
$g(p)$ would have an observable impact of \Lya\ data is simply a matter 
of assessing whether or not $\expt{v}$ exceeds 
the lower limit of this range.

By contrast, for Class-II velocity distributions, the situation is more 
complicated.  For such distributions, $\expt{v}$ no longer provides a reliable
characterization of the effect of free-streaming.  Indeed, for such 
distributions, the whole notion of a single ``free-streaming scale'' $\kFSH$ 
is inappropriate, given that different parts of the distribution free-stream 
at significantly different values of $k$.  For Class-II distributions, then, 
all parts of the phase-space distribution that lie within or above 
the range \il{10^{-8} \lesssim \expt{v} \lesssim 10^{-6}} can in principle 
affect \Lya\ data.  The corresponding constraints on such distributions are
therefore more subtle and must be dealt with more carefully.

\subsection{Applying \Lya\ Constraints to $T(k)$\label{subsec:Connecting}}

In order to determine the constraints on the transfer function $T(k)$ 
from \Lya\ data, knowledge of the spatial distribution of neutral 
hydrogen is necessary.  Calculating this distribution typically requires 
detailed hydrodynamic simulations, often in conjunction with $N$-body 
simulations.  Unfortunately, performing these numerical calculations is not 
always computationally feasible.  For this reason, it is useful to 
reformulate or ``recast'' these constraints in different approximate 
forms that may be easily applied to different candidate models of dark-matter 
physics.  

In general, we may group the recast methods that are in common use 
in the literature into two different classes. 
The first class comprises recasts which focus exclusively on the 
behavior of the power spectrum at relatively small values of $k$, 
while the second comprises recasts which incorporate the weighted 
behavior of the power spectrum over a broad range of wavenumbers,
typically including contributions from relatively large values of $k$.

In order to perform this analysis, we shall employ a single 
representative recast method from each class.
Our representative of the first class will be a well-known recast
method based on the so-called ``half-mode'' mass~\cite{Konig:2016dzg}.  
By contrast, our representative of the second class
will be a method based on the so-called $\dA$ estimator~\cite{Murgia:2017lwo}.
In this section, we shall review both of these methods.

\subsubsection{Analysis Based on the Half-Mode\label{subsubsec:halfmode}}

This method of assessing \Lya\ constraints on a given dark-matter model
exploits the fact that the sorts of numerical simulations described above 
have actually been performed~\cite{Bode:2000gq,Hansen:2001zv,Viel:2005qj}
for the case of one particular model of dark-matter physics:   
a model in which all of the dark matter is warm, with mass  $\mWDM$.  For 
such a model, the transfer function takes the approximate form
\beq
  \TWDM^2(k) ~\approx~ \left[1 + (\alpha k)^{2\nu}\right]^{-10/\nu} \ ,
  \label{eq:T2WDM}
\eeq
where \il{\nu = 1.12} and where 
\beq
\alpha \,=\, 0.049\left(\frac{\mWDM}{\rm keV}\right)^{\!\!-1.11} 
         \hskip -0.04truein 
               \left(\frac{\Omega_{\rm WDM}}{0.25}\right)^{\!\!0.11}
         \hskip -0.04truein 
             \left(\frac{h}{0.7}\right)^{\!\!1.22}\frac{\rm Mpc}{h} \,.
\eeq
We thus see that taking the dark matter to be warm rather than cold 
suppresses structure formation for wavenumber scales 
$k\gsim \calO(1/\alpha)$, but has essentially no effect for scales 
$k\ll 1/\alpha$, where $T(k)\approx 1$.  The scale $1/\alpha$ which 
characterizes the onset of suppression depends on the mass $\mWDM$ of 
the WDM particle, with smaller masses resulting in larger deviations 
from the predictions of CDM.~ 
\Lya\ constraints on this model then take the form of a lower bound on 
$\mWDM$, with the lower bounds $\mWDM\geq  3.5~{\rm keV}$ and  
$\mWDM\geq  5.3~{\rm keV}$ often quoted in the 
literature~\cite{Irsic:2017ixq}.  These two 
benchmark values correspond to different assumptions concerning the 
thermal history of the universe, and can thus be viewed as providing 
a measure of the uncertainty in the \Lya\ constraint.  

While these results apply to WDM, it is possible to exploit them in order 
to place constraints on other, more general models of dark matter.
Essentially the above WDM analysis can be viewed as imposing a lower limit 
on the transfer function $T(k)$ that can be consistent with \Lya\ constraints 
within the critical region of $k$-space which describes the onset of structure 
suppression.  To describe this onset region, one typically defines the so-called
``half-mode'' scale $\khalf$, which is simply the scale at which the 
squared transfer function has fallen to $1/2$:  
\beq
    \TWDM^2(\khalf) ~\equiv~ \frac{1}{2} \ .
\label{eq:khalfdef}
\eeq
One then takes the onset region to be that in which $k\lsim \khalf$.
Given Eq.~(\ref{eq:T2WDM}), we find that
\beq
    \khalf~\approx~ \frac{1}{\alpha}\left(2^{\nu/10} - 1\right)^{\!1/2\nu} \ .
\label{eq:khalf}
\eeq
For any general model of dark-matter physics with transfer function $T(k)$, 
we then ensure that we have satisfied \Lya\ constraints by demanding that 
\beq
     T(k)\geq \TWDM(k) ~~~~~{\rm  for~all}~ 0 \leq k\leq \khalf \ ,
\label{eq:halfmodeconstraint}
\eeq
where $\mWDM$ is set to its lower limit (either $3.5$~keV or $5.3$~keV).
Taking the larger critical WDM mass clearly leads to a more
stringent constraint.

Of course, a violation of the $T(k)\geq \TWDM(k)$ bound for {\it any}\/
value of $k$ can technically be considered a violation of the \Lya\ constraint.
However, this would be overly conservative, since  
a violation of the bound at very large $k$ is not
as critical as a violation at small $k$.  
It is for this reason that one demands $T(k)\geq \TWDM(k)$ only for
$k\leq \khalf$.

\subsubsection{Analysis Based on $\dA$\label{subsubsec:dA}}

An alternative approach to imposing \Lya\ constraints was proposed in
Ref.~\cite{Murgia:2017lwo}, building on results in Ref.~\cite{Schneider:2016uqi}.
Given a model of dark-matter physics and its associated matter power spectrum 
$P(k)$, we first calculate the corresponding ``one-dimensional'' power spectrum
\beq
    P_{\rm 1D}(k) ~\equiv~ \frac{1}{2\pi}\int_{k}^{\infty}dk' \, k'\, P(k') ~,
\eeq
which is simply the $k'$-integral of the matter power spectrum over all $k' > k$. 
We then define
\beq
  A ~\equiv~ \int_{\kmin}^{\kmax} dk 
    \frac{P_{\rm 1D}(k)}{P_{\rm 1D}^{\rm CDM}(k)} \ ,
\label{eq:Adef}
\eeq
where \il{\kmin = 0.5 \, h/{\rm Mpc}} and \il{\kmax = 20 \, h/{\rm Mpc}}, 
corresponding to the limits used in the \texttt{MIKE/HIRES} and \texttt{XQ-100} 
combined dataset~\cite{Irsic:2017ixq}.  We emphasize that $A$ is sensitive
to the behavior of $P(k)$ at all wavenumbers $k > \kmin$, including those 
above $\kmax$.

Given the definition in Eq.~(\ref{eq:Adef}), we immediately see 
that \il{A_{\rm CDM} = \kmax - \kmin}.  Our interest concerns the degree to 
which $A$ for a given dark-matter model differs from $A_{\rm CDM}$.
We therefore define the fractional deviation
\beq
    \dA \,\equiv\, \frac{A_{\rm CDM} - A}{A_{\rm CDM}}
    \,=\, 1 - \expt{\frac{\int_k^{\infty}dk'k' P(k')}{\int_k^{\infty}
    dk'k'P_{\rm CDM}(k')}}_{\!\! k}~,
\label{eq:dA}
\eeq
where $\langle\cdots\rangle_k$ indicates an
average over the range of comoving wavenumbers 
\il{\kmin \leq k \leq \kmax}.
We can then recast the \Lya\ constraints as placing a bound
\beq
     \dA ~\leq~ \dA_{\rm ref}~,
\label{dAlimit}
\eeq
where $\dA_{\rm ref}$ is a reference value obtained by applying 
Eq.~(\ref{eq:dA}) to a WDM model with $\mWDM$ set to its lower limit.  
Thus, once again, taking the larger critical WDM mass 
leads to a more stringent constraint.

It turns out that there is some disagreement in the literature concerning the precise numerical values for $\delta A_{\rm ref}$ that should be applied in Eq.~(\ref{dAlimit}).  
However,  we shall ultimately evaluate these quantities directly 
rather than rely on previously quoted results.  
The method by which we do this will be discussed further in Sect.~\ref{sec:recasting}.


\section{Recasting the Recasts: A ``Floating'' ${m_{\rm WDM}}$\label{sec:recasting}}


As we have seen, the analysis above is typically meant 
to produce a binary yes/no answer to the question of whether
a given dark-matter model is consistent with the \Lya\ constraints.
For example, in the half-mode analysis, a dark-matter model
is consistent with the \Lya\ constraints if and only if its transfer function 
$T^2(k)$ exceeds $\TWDM^2(k)$ for all $0\leq k\leq \khalf$,
in which $\TWDM^2(k)$ takes some reference value $\mWDM$
that is determined from observational data.

While a binary yes/no answer is suitable for assessing the viability 
of an isolated dark-matter model, we would like to explore the space 
of possible dark-matter models more generally.  That is, we would 
like to understand in a quantitative way how ``close'' to being ruled 
out a given model might be and whether a fine-tuning might be involved.
We also would like to understand how close a given model might come 
to being ruled out in the future, assuming further observational data, 
and we would also like to map out how these results might vary as we 
change the parameters in our underlying dark-matter model.  Finally, 
we would also like to {\it compare}\/ these two \Lya\ recasts in a 
meaningful and quantitative way.

In order to accomplish this, we look to extract more than a simple binary yes/no
viability decision from these recasts.  Instead, we would like to establish 
a ``viability parameter'' for each recast---a parameter which takes continuous values
and thereby enables quantitative studies of the sorts outlined above.
If formulated correctly, such viability parameters would also permit a direct comparisons 
between these different recasts. 

For the $\dA$ recast, it is natural to take the value of $\dA$ itself as
such a continuous viability parameter.  However, it is less obvious what might 
serve as a corresponding parameter for the half-mode recast.   One possibility, 
for example, might be to define a parameter $k_\times$
which corresponds to the wavenumber at which the transfer function $T^2(k)$ 
of a given dark-matter model crosses below the corresponding $\TWDM^2(k)$
in Eq.~\eqref{eq:T2WDM}.  We would then assess the viability 
of a given dark-matter model in terms of the gap 
between $k_\times$ and the half-mode $\khalf$.

Unfortunately, such viability parameters are unsuited for our purposes.
First, $k_\times$ and $\dA$ are completely different in their formulations 
and cannot be directly compared.  Indeed, they do not even have the same units.
Moreover, the definition of $k_\times$ intrinsically depends on 
the particular reference value $\mWDM$.
Thus, any changes in this reference value (such as might occur in the future
as further observational data is collected)
would change the value of $k_\times$ associated with 
a given dark-matter model, even if the model itself is unchanged.

For such reasons, we shall now formulate two different parameters
by exploiting the single ``common denominator''
that underlies both of these \Lya\ recasts, 
namely the WDM model for which full \Lya\ simulations have been performed.
Clearly, the predictions of a given WDM model depend on the 
mass $\mWDM$ of the supposed dark-matter candidate.
We shall therefore exploit this observation by 
using each recast method in order to determine 
{\it the effective value of $\mWDM$ which would place the given model
directly on the critical allowed/disallowed boundary}\/.
These effective values will then be identified 
as $\mhalf$ and $\mdA$, respectively.

More specifically, we shall define $\mhalf$ as follows.
For any proposed model of dark-matter physics, there exists
a corresponding transfer function $T^2(k)$.  We identify $\mhalf$ as the 
maximum value of $\mWDM$ for which $T^2(k)\geq \TWDM^2(k)$ for all 
$0\leq k\leq \khalf$, where $\TWDM^2(k)$ is given in terms of $\mWDM$ through 
Eq.~(\ref{eq:T2WDM}).  In this way, $\mhalf$ is identified as the critical 
value of $\mWDM$ for which our proposed dark-matter model would have resided 
directly on the critical line between being allowed and disallowed, according 
to the half-mode recast.  Of course, a given dark-matter model itself may 
contain various adjustable parameters as part of its definition.  This 
critical value $\mhalf$ of $\mWDM$ will then ``float'' as we move across
the parameter space of the model.

\begin{figure}[t]
    \begin{center}
    \includegraphics[width=0.49\textwidth, keepaspectratio]{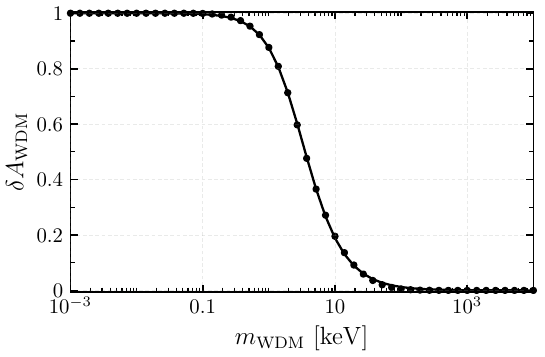}
        \caption{
            Relation between $\dAWDM$ and $\mWDM$.  
            The point markers correspond to numerical results 
            obtained through Eq.~(\ref{eq:prefit}), while the solid 
            curve corresponds to the best-fit parametrization 
            in Eq.~\eqref{eq:dAWDMfit}.
            }
            \label{fig:dAmWDMmapping}
    \end{center}
\end{figure}

We shall follow a similar procedure in defining $\mdA$.
At first glance, defining  $\mdA$ is not as straightforward, 
since there is no general map between $\dA$ and $\mWDM$ 
given in the literature.  
Indeed, only a few particular values of $(\mWDM,\dA)$ are quoted, 
usually corresponding to the
more and less conservative 
\Lya\ bounds on $\mWDM$.  
Nevertheless, we can build such a map directly. 
Starting from $\TWDM^2(k)$ in Eq.~\eqref{eq:T2WDM},
we can calculate the corresponding one-dimensional WDM spectrum
\beq
       \left[P_{\rm WDM}\right]_{\rm 1D} ~=~ \frac{1}{2\pi}
       \int_k^{\infty}dk'\, k'\, \TWDM^2(k')P_{\rm CDM}(k') ~
\label{eq:prefit}
\eeq
where we use \texttt{CLASS} to evaluate $P_{\rm CDM}(k')$.
Using Eq.~(\ref{eq:dA}), we can then calculate $\dAWDM$.
Carrying out this procedure numerically, we obtain
the results shown in Fig.~\ref{fig:dAmWDMmapping}.  
In this figure, the discrete point markers indicate our numerical results.
We find that these numerical results follow the functional form
\beq
     \dAWDM ~=~ \left[1 + a\left(\frac{\mWDM}{\rm keV}\right)^{b}\right]^{-c}
\label{eq:dAWDMfit}
\eeq
with best-fit parameters 
\il{\lbrace a,b,c\rbrace = \lbrace 0.19, 1.63, 0.75\rbrace}.
As is apparent from Fig.~\ref{fig:dAmWDMmapping}, our numerical results are 
fit remarkably well by this function across the four orders of magnitude 
in $\mWDM$ most relevant for our purposes.  

Equipped with a map between $\dAWDM$ and $\mWDM$,
we may now identify $\mdA$ for a given dark-matter model
as the value of $\mWDM$ for which $\dA= \dAWDM$.
Equivalently, given the value of $\dA$ for our model,
we can simply invert Eq.~\eqref{eq:dAWDMfit} to define
\beq
    \mdA ~\equiv~ \left[\frac{(\dA)^{-1/c} - 1}{a}\right]^{1/b} 
     ~{\rm keV} ~.
\eeq

For any proposed model of dark-matter physics,
we now have a recipe for calculating
two mass scales, $\mhalf$ and $\mdA$, 
corresponding to the two different recast methods.
Each of these is a continuous parameter which varies across the 
parameter space of a given dark-matter model and which quantifies 
how close to satisfying the \Lya\ bounds 
the different recast methods find the model to be.
In other words, $\mhalf$ and $\mdA$ independently assess 
the {\it viability}\/ of the proposed dark-matter model.

However, these variables not only permit us to construct \Lya\ 
constraints --- they also permit us to {\it compare}\/ our two recast 
methods in a quantitative manner by evaluating the quotient
\beq
    R ~\equiv~ \frac{\mhalf}{\mdA} ~.
\label{eq:R}
\eeq
Indeed, just like $\mhalf$ and $\mdA$, this comparator $R$ will 
generally also vary across the parameter space of a given dark-matter model.
On the one hand, when $R\approx 1$, the two recasts are consistent 
with each other and provide similar results.
On the other hand, when $R$ differs significantly from $1$, the 
recasts are inconsistent with each other.  In this case, the choice 
of recast has a significant effect on the corresponding \Lya\ constraints,
producing different exclusion regions for the same dark-matter model.

The questions we face, then, are two-fold.  First, we would like to 
determine the sorts of general dark-matter models and associated 
parameter-space regions which are generally viable or excluded by \Lya\ 
constraints.  At the same time, however, we also seek to determine for 
which classes of dark-matter models --- and for which portions of their 
associated parameter spaces --- we obtain $R \approx 1$, and for which 
classes of models and regions of associated parameter space we do not. 
It is to these questions that we now turn.

\section{\Lya\ Constraints on Class-I Velocity Distributions\label{sec:minimal}}


Our aim in this paper is to investigate the \Lya\ constraints
on a multitude of possible dark-matter velocity distributions.  
In conducting this investigation, we shall proceed systematically, 
beginning with the simplest such distributions and then broadening the 
scope of our analysis to include more complicated, multi-modal ones.
In this section, we shall begin by focusing on Class-I velocity distributions, 
including thermal distributions and distributions which can be approximated
as Gaussian.  We shall then expand our analysis to consider Class-II 
velocity distributions in Sect.~\ref{sec:nonminimal}.

Before we proceed, however, we emphasize once again that a lack of 
complexity in the dark-matter velocity distribution does not necessarily 
reflect a corresponding lack of complexity in the particle content of 
the dark sector or in the interaction Lagrangian which governs the 
dynamics within that sector.  Indeed, one can easily imagine complicated 
dark sectors which involve large numbers of particle species and/or 
non-trivial interactions, but which nevertheless yield dark-matter 
velocity distributions which are trivial in terms of their overall
shapes.

\subsection{Thermal Distributions\label{subsec:thermal}}

In order to establish a baseline, we shall first consider velocity 
distributions similar to the canonical distribution which arises in WDM 
scenarios.  Indeed, such velocity distributions can be expected to arise 
for dark-matter particles of mass $m$ which are initially in thermal 
equilibrium  with a radiation bath, but subsequently decouple from 
that bath at a time prior to $\tMRE$ when the bath temperature $T_d$ 
greatly exceeds $m$.  The present-day phase-space distribution for such 
a dark-matter species is then given by the Maxwell-Boltzmann form
\begin{align}
    g(p) ~&=~  \frac{1}{2}\,\calN \left(\frac{p}{\Td\ad}\right)^{\!\!3} 
    \exp\left[-p/(\Td\ad)\right] \ ,
\label{eq:gthermal}
\end{align}
where we have ignored modifications due to Fermi/Bose statistics,
where $\calN$ is a normalization factor defined according to 
Eq.~\eqref{eq:Ndef}, and where $\ad$ denotes the value of the scale factor 
at $t_d$.  The average momentum for a population of dark-matter particles 
with a phase-space distribution given by Eq.~\eqref{eq:gthermal} is
\beq
    \expt{p} ~=~ 3\Td\ad \ .
    \label{eq:thermalaverage}
\eeq
This result implies that the $g(p)$ distribution in Eq.~\eqref{eq:gthermal} 
is completely specified by $\expt{p}$ and can be viewed as a function of 
the ratio \il{p/\hspace{-1mm}\expt{p}}.

It is well known that $\expt{p}$ 
plays a crucial role in determining how cosmological perturbations
grow in scenarios with $g(p)$ distributions of this sort.  Indeed,
as indicated in Eq.~\eqref{eq:dFSH}, $\expt{v} \approx \expt{p}/m$ 
effectively sets the free-streaming horizon for the dark 
matter and thereby the wavenumber $\kFSH$ above which $P(k)$ is suppressed.
However, the \emph{shape} of the distribution can also 
be important~\cite{Dienes:2020bmn,Dienes:2021itb}, particularly if the higher moments
of the distribution --- such as the standard deviation or the 
skewness --- are large.  

For the phase-space distribution in Eq.~\eqref{eq:gthermal}, we observe 
that these higher moments are all independent of $\expt{p}$.  Indeed, 
this is related to the fact that a change in the value 
of $\expt{p}$ does not affect the {\it shape}\/ of this distribution
in $\log p$-space, but instead merely shifts it uniformly to higher
or lower values of $\log p$.  For example, the standard deviation of 
$\log p$ for this distribution turns out to be 
\begin{align}
    \sigma ~&\equiv~ \sqrt{\langle 
    (\log p - \langle \log p\rangle)^2\rangle} \nn \\
    ~&=~ \sqrt{\frac{\pi^2}{6} - \frac{5}{4}} ~\approx~ 0.63 \ .
\label{eq:thermalwidth}
\end{align}
Likewise, the skewness for this distribution is
\begin{align}
    S ~&\equiv~ \frac{\langle (\log p - \langle\log p\rangle)^3\rangle}
    {\sigma^{3}} \nn \\
    ~&=~ \frac{\frac{9}{4} - 2\zeta(3)}{\left(\frac{\pi^2}{6} 
    - \frac{5}{4}\right)^{3/2}} ~\approx~ -0.62 \ .
    \label{eq:thermalskew}
\end{align}
The negative value of $S$ indicates that the distribution ``tilts'' to 
the right as compared with one which would have been perfectly symmetric 
in $\log\,p$-space.

\begin{figure}[t]
    \begin{center}
    \includegraphics[width=0.49\textwidth, keepaspectratio]{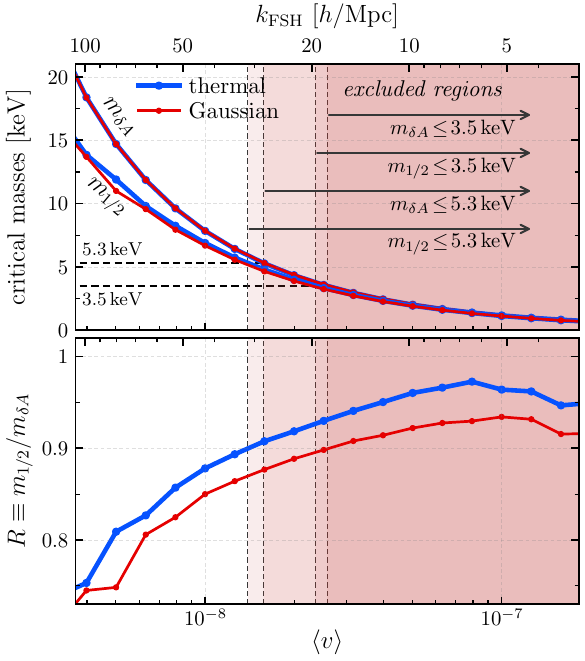}
        \caption{
            The critical masses $\mhalf$ and $\mdA$ (upper panel) 
            and the comparator $R$ (lower panel), plotted as functions of 
            $\expt{v}$ for the thermal velocity distribution in 
            Eq.~\eqref{eq:gthermal} and the Gaussian velocity distribution 
            to be discussed in Eq.~\eqref{eq:glognormal}, the latter with a 
            width $\sigma= 0.63$ taken to match that of the thermal 
            distribution.  Also shown are the \Lya\ exclusion regions for 
            the thermal velocity distribution according to our half-mode 
            and $\dA$ recasts for the critical masses $3.5$~keV and 
            $5.3$~keV.  The top axis indicates the dark-matter free-streaming 
            horizon wavenumber $k_{\rm FSH}$ corresponding to each value of 
            $\expt{v}$. We observe that within the observationally 
            viable regions of $\expt{v}$, the comparator $R$ departs 
            considerably from unity, indicating a meaningful disagreement 
            between the results of the two recasts.
            }
            \label{fig:thermal}
    \end{center}
\end{figure}

We now seek to assess for which velocities $\expt{v}$ the \Lya\ 
constraints can be satisfied, given the distribution in 
Eq.~\eqref{eq:gthermal}.  In order to do this, we evaluate the 
transfer function $T^2(k)$ which corresponds to the dark-matter 
distribution $g(p)$ for many different values of $\expt{v}$. 
We then employ the recasts from Sect.~\ref{sec:recasting} 
in order to obtain information about the \Lya\ bounds.

In the upper panel of Fig.~\ref{fig:thermal}, we plot the critical
masses $\mhalf$ and $\mdA$ (blue curves) for the dark-matter
phase-space distribution in Eq.~\eqref{eq:gthermal} as functions of 
$\expt{v}$.  In the lower panel of this same figure, we plot 
the corresponding value of the comparator \il{R\equiv \mhalf/\mdA} 
(red curve).  Since the relation between $\expt{v}$ and the free-streaming 
horizon in Eq.~\eqref{eq:dFSH} establishes a one-to-one correspondence between
values of $\expt{v}$ and values of the comoving wavenumber $\kFSH$ for 
Class-I velocity distributions such as this one, tick marks indicating
the values $\kFSH$ have also been included at the top of each panel. 

We immediately see from Fig.~\ref{fig:thermal} that while $\mhalf$
and $\mdA$ agree quite well for large $\expt{v}$, a significant
difference between them begins to develop as $\expt{v}$ decreases.
As a result of this difference, the constraints on $\expt{v}$ obtained
from each of the two recasts differ.  Indeed,   
when we employ the less conservative bound \il{\mWDM \gtrsim 3.5\,{\rm keV}}
in implementing each recast, we find that the half-mode analysis excludes the 
light-gray region on the right side of the figure, while the $\dA$ analysis 
also excludes the thin, dark-gray stripe.  Likewise, if we employ the more
conservative bound \il{\mWDM \gtrsim 5.3\,{\rm keV}}, 
we find that the half-mode analysis further excludes the dark-red region,
while the $dA$ analysis also excludes the thin, light-red stripe.  

While these differences are not terribly significant, we observe that the 
difference between $\mhalf$ and $\mdA$ becomes more and more pronounced
as $\expt{v}$ decreases.  Indeed, we observe that the comparator falls to 
\il{R\approx 0.72} at the left edge of Fig.~\ref{fig:thermal}.  
That such a difference between recast predictions can arise even 
for the case of a simple, Class-I velocity distribution suggests that 
significantly smaller $R$ values could arise for other, more complicated
distributions.  As we shall see, this indeed turns out to be the case.

\subsection{Gaussian Distributions\label{subsec:gaussian}}

The thermal dark-matter phase-space distribution in Eq.~\eqref{eq:gthermal} 
provides a well-motivated and simple baseline from which to begin
our exploration of more general functional forms of $g(p)$.
However, since the parameters characterizing the shape of this baseline 
distribution --- \ie, its width $\sigma$, its skewness $S$, \etc --- were 
completely specified, our analysis thus far has exposed the sensitivity of 
the \Lya\ constraints to a single variable only, namely the average 
velocity $\expt{v}$.  

\begin{figure*}[t]
    \begin{center}
    \includegraphics[width=1.0\textwidth, keepaspectratio]{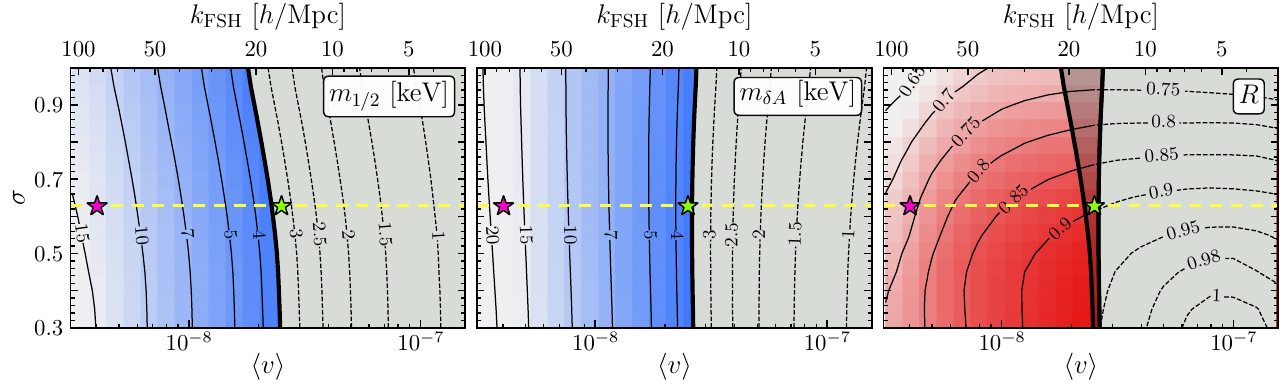}
        \caption{Contours in the $(\expt{v},\sigma)$-plane of the critical 
        masses $\mhalf$ (left panel) and $\mdA$ (center panel), as 
        well as the comparator $R$ (right panel), for the Gaussian 
        phase-space distribution in Eq.~\eqref{eq:glognormal}.  
        The thick black curves in the left and center panels indicate 
        the \Lya\ constraint obtained by employing the less conservative 
        bound \il{\mWDM \gtrsim 3.5\,{\rm keV}} when implementing 
        the corresponding recast.  The gray region to the right of 
        each thick black curve is excluded.  The exclusion contours for both
        of these recasts are also superimposed on the contours of $R$ 
        shown in the right panel.  The dashed yellow line in each panel 
        corresponds to \il{\sigma = 0.63}, which matches the standard 
        deviation of the distribution in Eq.~\eqref{eq:gthermal}.
        The magenta and green stars indicate the locations in the 
        $(\expt{v},\sigma)$-plane for which the corresponding transfer 
        functions are plotted in Fig.~\protect\ref{fig:unimodalPk}.
        }
        \label{fig:unimodal}
    \end{center}
\end{figure*}

We would now like to extend our study and examine how variations 
in the {\it shape}\/ of a unimodal distribution modify the \Lya\ constraints.
To this end, we shall now consider the case in which $g(p)$ takes the form 
of a Gaussian distribution in $(\log\,p)$-space:
\begin{align}
  g(p) ~&=~ \frac{\calN}{\sqrt{2\pi \sigma^2}} 
    \exp\left[ - \frac{ (\log p - \langle \log p\rangle)^2}{2\sigma^2} 
    \right] \ ,
    \label{eq:glognormal}
\end{align}
where we take the mean $\langle \log p\rangle$ and standard deviation 
$\sigma$ to be free parameters, and where $\calN$ is a normalization 
factor.  We note that for a log-normal distribution of this sort we may
also express $\expt{\log p}$ in terms of $\expt{p}$ --- or, equivalently,
in terms of \il{\expt{v}\approx \expt{p}\!/m} --- through the relation
\il{\langle \log p\rangle = \log\langle p\rangle - \sigma^2/2}.  This
relation will allow us to compare the results we obtain for this 
phase-space distribution directly with those obtained from the 
baseline thermal distribution in Eq.~\eqref{eq:gthermal}. 

We shall begin our analysis of the phase-space distribution in 
Eq.~\eqref{eq:glognormal} by considering how the difference in
shape between this distribution and the baseline thermal 
distribution in Eq.~\eqref{eq:gthermal} affects our results for 
$\mhalf$ and $\mdA$.  In order to facilitate a direct comparison
between these two distributions, we fix the standard deviation 
for our Gaussian to the value \il{\sigma = 0.63} which 
accords with Eq.~\eqref{eq:thermalwidth} and examine how these two
critical masses behave as functions of $\expt{v}$.  Our results
for $\mhalf$ and $\mdA$ are indicated by the red curves in the 
upper panel of Fig.~\ref{fig:thermal}, while the corresponding
results for $R$ are indicated by the red curve in the lower panel.
The $\mdA$ curve for the Gaussian distribution coincides almost exactly 
with the corresponding curve for the baseline thermal distribution in 
Eq.~\eqref{eq:gthermal} --- so much so, in fact, that it is effectively 
hidden beneath that curve.  By contrast, the $\mhalf$ curve for the
Gaussian distribution departs appreciably from the corresponding
curve for the baseline distribution, and the value of $R$ is 
consequently smaller for the Gaussian.  These results indicate that
even for unimodal, Class-I velocity distributions, the shape of 
the distribution can have a small but appreciable impact on the results 
obtained from the half-mode recast.  Nevertheless, we do not find a 
particularly large difference between the \Lya\ bounds obtained for 
these two distributions.

We now expand our analysis of the phase-space distribution in 
Eq.~\eqref{eq:glognormal} in order to examine the effect of varying 
$\sigma$ as well as $\expt{v}$. 
Indeed, by varying $\sigma$ as well as $\expt{v}$ , we are now actually 
varying the shape of the log-normal distribution in Eq.~\eqref{eq:glognormal} 
and not merely shifting this distribution rigidly along the $(\log\,p)$-axis.
In Fig.~\ref{fig:unimodal}, we display contours of $\mhalf$ (left panel) 
and $\mdA$ (center panel) in $(\expt{v},\sigma)$-space for the velocity 
distribution in Eq.~\eqref{eq:glognormal}.  The thick black curve in each of
these two panels panel shows the \Lya\ bound obtained by employing the 
less conservative bound \il{\mWDM \gtrsim 3.5\,{\rm keV}} in implementing 
the corresponding recast.  The gray region to the right of each each thick
black curve is excluded.  In the right panel of the figure, the exclusion 
contours for both of these recasts are superimposed on the contours of 
the comparator $R$.  For $\sigma=0.63$ (yellow line shown in each panel),  
the results obtained for $\mhalf$ and $\mdA$ 
correspond to those plotted in Fig.~\ref{fig:thermal}.

Once again, we observe that while there are regions of parameter space
wherein the results of the half-mode and $\dA$ recasts coincide quite well,
there are other regions --- especially those within which $\expt{v}$ is 
small or $\sigma$ is large --- wherein $R$ differs significantly from unity.
The results shown in the left panel indicate that the constraint derived
from the half-mode recast tightens by roughly $30\%$ over the region of 
parameter space shown as $\sigma$ increases and $g(p)$ becomes broader.  
Ultimately, this stems from the fact that a broader phase-space distribution 
includes a non-negligible population of dark-matter particles at higher 
velocities that can smooth out structure on larger scales.  As a result, 
$\mhalf$ decreases as $\sigma$ increases, leading to more stringent 
constraints on models with broader $g(p)$ distributions.
By contrast, the results shown in the center panel of Fig.~\ref{fig:unimodal} 
indicate that the constraint derived from the $\mdA$ recast does not 
depend sensitively on the value of $\sigma$.  Indeed, across the range of
$\sigma$ shown, this constraint is approximately the same as that obtained
for our baseline thermal distribution in Eq.~\eqref{eq:gthermal}.

The results for $R$ displayed in the right panel  of Fig.~\ref{fig:unimodal} 
indicate that the half-mode and $\dA$ recasts yield similar bounds when 
$\sigma$ is relatively small.  However, for \il{\sigma \gtrsim 0.5}, 
the constraints obtained from the two recasts diverge, with $R$ deviating
significantly from unity across the upper portions of the panel.  
Furthermore, the results for $R$ within the region of 
$(\expt{v},\sigma)$-space which is not currently excluded by \Lya\ data
indicate how the \Lya\ constraints would change if future observations 
were to further tighten the \Lya\ bound on $\mWDM$.  In particular, we 
find that the disagreement between recasts becomes more severe as 
$\expt{v}$ decreases and as $\sigma$ increases.

\begin{figure}[t]
    \begin{center}
    \includegraphics[width=0.49\textwidth, keepaspectratio]{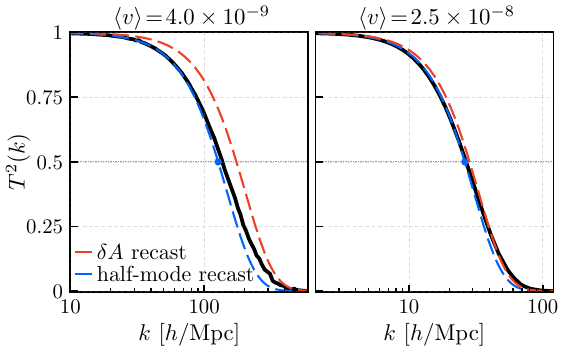}
        \caption{
            The transfer functions $T^2(k)$ 
            which correspond to the parameter-space points indicated by the 
            magenta and green stars in Fig.~\ref{fig:unimodal}.  The results
            in the left panel correspond to the magenta star, which is situated
            within a region of parameter space which lies far from the 
            exclusion region and for which
            the value of $R$ is relatively small, while the results
            in the right panel correspond to the green star, which is 
            situated near the \Lya\ exclusion contours associated with
            both recasts.  The solid black curve
            in each panel represents the result of a numerical calculation, 
            while the dashed curves represent the 
            WDM transfer functions onto which this 
            $T^2(k)$ is effectively mapped by the half-mode and $\dA$ recasts. 
            }
            \label{fig:unimodalPk}
    \end{center}
\end{figure}

The results shown in Fig.~\ref{fig:unimodal} beg the question whether 
an even more dramatic departure from the results obtained for our 
baseline thermal distribution in Eq.~\eqref{eq:gthermal} would be obtained 
for Gaussian $g(p)$ distributions with even larger values of $\sigma$.  
Unfortunately, within this region of parameter space --- particularly for 
small values of $\expt{v}$ --- assessing the values of $\mhalf$ and $\mdA$ 
becomes challenging in practice, as the simulations required to evaluate 
$T^2(k)$ become significantly more computationally expensive.  
Nevertheless, based on our qualitative understanding of how the bounds 
obtained from the $\mhalf$ and $\mdA$ recasts arise, we would expect that 
the bounds obtained from both recasts become significantly tighter when 
$\sigma$ becomes increasingly large.

Further insight into the results shown in Fig.~\ref{fig:unimodal} can
be gleaned from examining how the shape of the transfer function itself 
varies across $(\expt{v},\sigma)$-space.  In Fig.~\ref{fig:unimodalPk},
we show the numerical results for $T^2(k)$ (solid black curves) for two 
different points within that parameter space.
The curve in the left panel corresponds to the point 
$(\expt{v},\sigma) = (4\times 10^{-9},0.63)$ indicated by the magenta star 
in Fig.~\ref{fig:unimodal}, while the curve in the right panel 
corresponds to the point $(\expt{v},\sigma) = (2.5\times 10^{-8},0.63)$
indicated by the green star.  In
each panel, alongside the true $T^2(k)$ curve obtained for the Gaussian
$g(p)$ distribution in question, we also plot the $T^2(k)$ curves 
obtained for WDM models with $\mWDM$ set equal to the value of $\mhalf$ 
(dashed blue curve) and $\mdA$ (dashed red curve) that we obtain for this
$g(p)$ distribution.  In other words, these dashed blue and red curves 
respectively represent the WDM transfer functions onto which the true
$T^2(k)$ is effectively mapped by the half-mode and $\dA$ recasts. 

We see from the right panel of Fig.~\ref{fig:unimodalPk} that the
$T^2(k)$ curves associated with both recasts nearly coincide both
with each other and with the true $T^2(k)$ curve obtained for this 
parameter-space point.  Indeed, this result is not unexpected, given 
that $R\approx 0.90$ is reasonably close to unity 
for this point, which lies very close to the \Lya\ exclusion contours in
Fig.~\ref{fig:unimodal} associated with both recasts.  By contrast, 
for the point shown in the 
left panel, not only do the WDM transfer functions associated with these 
two recasts differ appreciably from each other, they also both differ 
appreciably from the true $T^2(k)$ curve.  Again, this is not unexpected, 
given that $R\approx 0.74$ differs significantly from 
unity for this point.  However, the results shown in this panel
illustrate the ways in which these recasts can yield unreliable results
when the shape of the dark-matter velocity distribution differs
from that of WDM.~  Moreover, they also illustrate how sensitively 
the \Lya\ constraints depend on the precise form of $T^2(k)$.
Indeed, as we shall now see, these issues can become
even more pronounced for Class-II distributions.

\section{\Lya\ Constraints on Class-II Velocity Distributions\label{sec:nonminimal}}


We now expand our analysis to include a broader variety of dark-matter
velocity distributions --- distributions which depart more dramatically
from the simple, unimodal Class-I distributions we have considered thus
far.  One particularly interesting possibility along these lines is to 
consider distributions which are multi-modal.  Indeed, Class-II
distributions of this sort arise in a variety of dark-matter scenarios
in which non-negligible contributions to the dark-matter abundance
occur at different timescales.  Such scenarios include those in which 
freeze-out or freeze-in~\cite{Konig:2016dzg,Du:2021jcj,Decant:2021mhj} production is 
accompanied by production from the out-of-equilibrium decays of
heavy particles at late times; those in which different decay channels 
for the same unstable particle species with different decay kinematics 
contribute non-negligibly to the dark-matter abundance~\cite{Heeck:2017xbu};
and those in which dark-matter is produced non-thermally through decay 
cascades~\cite{Dienes:2020bmn}.

In order to examine multi-modal velocity distributions of this sort, 
we shall consider a phase-space distribution $g(p)$ which consists 
of a sum of Gaussian peaks, each of which takes the form specified in
Eq.~\eqref{eq:glognormal} and each of which provides a contribution 
$\Omega_i$ toward the total present-day dark-matter abundance 
\il{\Omega\equiv\sum_i\Omega_i \approx 0.26}~\cite{Planck:2018vyg}.
Such a form for $g(p)$ is indeed nothing but a Gaussian decomposition
of the phase-space distribution in $\log p$-space, and as such is
capable of approximating most distributions of interest.
We shall parametrize this phase-space distribution as follows:
\beq
  g(p) ~\approx~ \!\!\!\sum_{i=0}^{N_G-1}
    \!\!\frac{\mathcal{N}\Omega_i}{\sqrt{2\pi}\sigma_i\Omega}
    \!\exp\left\{\!-\frac{1}{2\sigma^2_i}
    \left[\log\left(\!\frac{p}{\pj{i}}\!\right) +
    \frac{1}{2}\sigma^2_i\right]^2\!\right\} \ ,
  \label{eq:gsum}
\eeq
where $\expt{p}_i$  is the average momentum of the $i^{\, \rm th}$ 
Gaussian, where $\sigma_i$ is the corresponding width, where $N_G$ is 
the number of Gaussian components included in the sum, and where 
the index \il{i = 0, 1, 2,\ldots, N_G-1} labels these 
components in order of {\it decreasing}\/ $\expt{p}_i$.

\begin{figure*}[t]
    \begin{center}
    \includegraphics[width=0.49\textwidth, keepaspectratio]{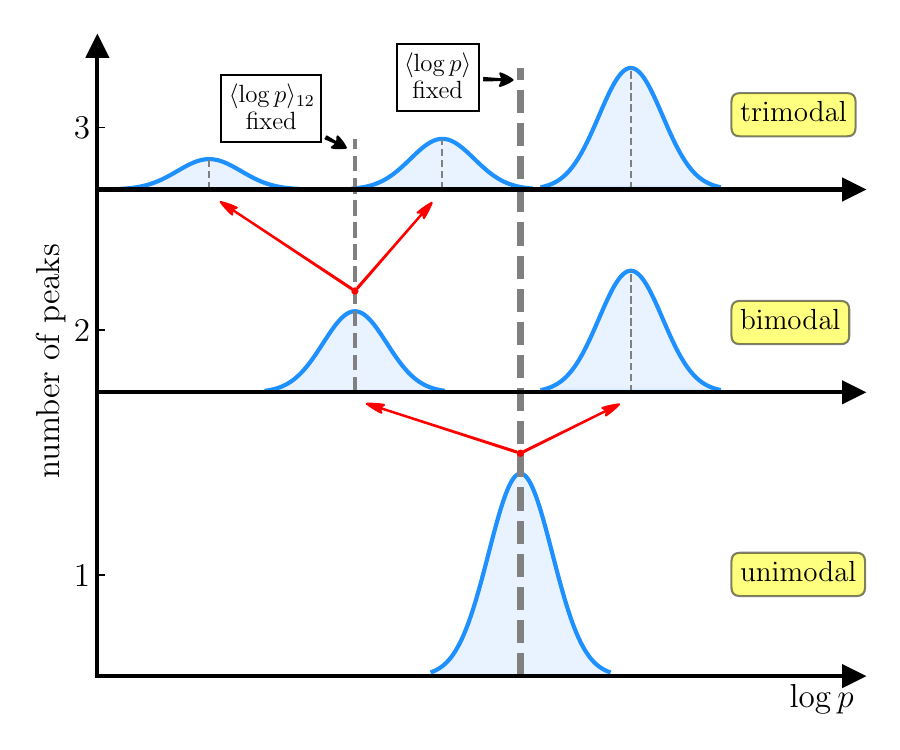}
        \includegraphics[width=0.49\textwidth, keepaspectratio]{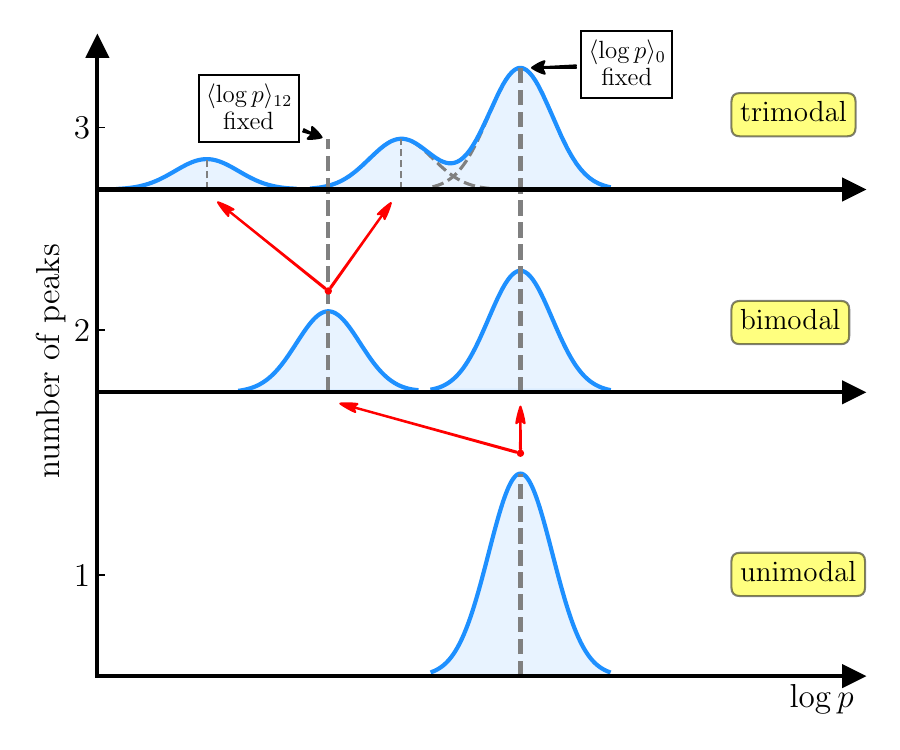}
        \caption{
          Two schematics illustrating methods through which multi-modal 
          dark-matter velocity distributions can be generated from a single 
          unimodal distribution.  {\it Left panel:}\/  The lowest-velocity 
          peak at each step is bifurcated into two peaks in such a manner as 
          to preserve the ``center of velocity'' $\left<\log p\right>_{ij}$
          defined in Eq.~(\protect\ref{eq:weightedavggeneric}) for
          the two-peak system, while keeping $\left<\log p\right>$ fixed. 
          {\it Right panel:}\/ Here the bifurcation procedure keeps both 
          $\left<\log p\right>_{ij}$ and $\left<\log p\right>_{0}$ fixed.  
          The analysis to be discussed in Sects.~\ref{subsec:twinpeaks} 
          and \ref{subsec:threepeaks} is based on this approach. 
          \label{fig:bifurcations}}
    \end{center}
\end{figure*}

Since a $g(p)$ distribution of this form with $N_G$ Gaussian components 
is described by $3N_G -1$ free parameters, systematically surveying the 
parameter space of possible $g(p)$ distributions rapidly becomes 
impractical as $N_G$ increases.  It will therefore be necessary for us 
to impose certain simplifying assumptions if we wish to 
explore the landscape of possible $g(p)$ distributions 
of the form given in Eq.~(\ref{eq:gsum}) in a meaningful yet tractable way.
One of our principal aims in this paper is to assess how the 
{\it detailed shape}\/ of a particular feature in $g(p)$ --- not merely the 
average velocity of the particles associated with that feature, but the 
full distribution of velocities --- affects the results
of the half-mode and $\dA$ recasts.  Whatever procedure we adopt 
should further this aim to whatever extent possible and highlight the 
ways in which the detailed shape of $g(p)$ impacts the \Lya\ constraints 
derived from the half-mode and $\dA$ recasts.  

An analysis procedure which furthers this aim is illustrated schematically 
in the left panel of Fig.~\ref{fig:bifurcations}.  According to this procedure, one begins 
with a $g(p)$ distribution consisting of a single Gaussian peak with a 
specified width $\sigma_0$ and
average velocity $\expt{v}_0$.  One then splits this peak 
into two individual peaks whose individual widths $\sigma_0$ and $\sigma_1$ 
are each equal to the width of the original peak and whose total
abundance $\Omega_0 + \Omega_1$ is equal to the abundance of the original 
peak.  Next, one separates these new peaks in velocity-space by endowing 
them with different average velocities $\expt{v}_0$  and $\expt{v}_1$.  
However, one does this in such a way that the abundance-weighted 
``center of velocity'' $\expt{\log v}_{01}$ for these peaks --- a
quantity which we define as
\begin{equation}
    \expt{\log v}_{ij} ~\equiv~ 
    \frac{\Omega_{i}}{\Omega_{i} + \Omega_{j}}\!
    \expt{\log v}_{i} 
    + \frac{\Omega_{j}}{\Omega_{i} + \Omega_{j}}\!
    \expt{\log v}_{j}
  \label{eq:weightedavggeneric}
\end{equation}
for any two peaks labeled by indices $i$ and $j$ ---
remains equal to the center of velocity of the original peak.  
One may then iterate this procedure by taking the lowest-velocity peak
in the resulting distribution, splitting it into two peaks whose
individual widths $\sigma_1$ and $\sigma_2$, total abundance 
$\Omega_1 + \Omega_2$, and center of velocity $\expt{\log v}_{12}$ 
are equal to the abundance and center of velocity of that original peak,
and so forth. In this procedure, $\left<\log p\right>$ remains fixed 
for the total population of dark-matter components. 

This iterative procedure has several features.
First, each time we perform such a bifurcation, we introduce only two additional
free parameters: one which characterizes the relative abundances of the
two resulting peaks and one which characterizes their separation in 
$(\log p)$-space.  Second, since the shape of the matter power spectrum is quite
sensitive to small modifications of $g(p)$ at large $p$, it is useful to
adopt a procedure in which one examines the effect of modifying the shape
of $g(p)$ of small $p$ while leaving the structure at large $p$ fixed. 
Third, since we hold the center of velocity fixed when we perform this 
bifurcation, any difference between the \Lya\ 
constraints that we obtain for the resulting $g(p)$ distribution relative
to the original distribution is solely a consequence of the differences in
their detailed shapes.

More mathematically, we can understand this approach as follows.  Like any 
distribution function, $g(p)$ can be described in terms of its {\it moments}. 
The zeroth moment of any distribution is the total area under the curve.  
For $g(p)$, this is the total dark-matter abundance, which we are holding fixed.  
The first moment of any distribution is then the average value of its independent 
variable --- in this case, the average velocity $\langle \log v\rangle_{ij}$.  
Holding this quantity fixed therefore enables us to disentangle the effects of 
these first two moments of $g(p)$ on the \Lya\ constraints, and thereby isolate 
the dependence of the \Lya\ constraints on the {\it higher}\/ moments of the 
distribution which concern the {\it distribution}\/ of the dark-matter 
velocities {\it around}\/ this average velocity.  Indeed, in some sense this 
has been the goal of our study all along, given that Class~I and Class~II 
distributions differ in precisely how these velocities are distributed 
relative to the mean.

This procedure, although mathematically elegant, is computationally difficult 
to realize in practice.  However, we may follow a slightly modified version 
of this procedure which is also capable of surveying phenomenologically 
rich portions of the multi-modal parameter space.  This alternative procedure
is shown in the right panel 
of Fig.~\ref{fig:bifurcations}.  As before, one begins with a $g(p)$ distribution 
consisting of a single Gaussian peak with a width $\sigma_0$ and average velocity 
$\expt{v}_0$, and splits it into two peaks.  However, instead of holding the 
center of velocity of the system fixed, one holds $\left<\log v\right>_0$ fixed 
and fission off some of the abundance of this peak in order to construct a second 
peak with average velocity \il{\left<\log v\right>_1 < \left<\log v\right>_0}.
One then proceeds by bifurcating this colder peak in the same way that one would 
have done it in the original procedure, by holding $\left<\log v\right>_{12}$ fixed, 
and so forth.  This modified procedure allows us examine the impact of the 
lower-velocity peak on the \Lya\ constraints for the \il{N_G = 2} case in a more
straightforward manner.  In what follows, we adopt the procedure outlined in the 
right panel in our analysis.

\subsection{Bi-Modal Distributions\label{subsec:twinpeaks}}

We begin our analysis of the \Lya\ constraints on phase-space 
distributions of the general form in Eq.~\eqref{eq:gsum} by 
considering the simplest non-trivial case --- the
case of distributions with \il{N_G = 2} Gaussian components.
For simplicity, in exploring the five-dimensional parameter 
space which characterizes these double-peak $g(p)$ distributions, 
we shall set the widths of the two Gaussian components to be equal 
to each other and focus on the effects of varying the average velocity 
$\expt{v}_0 \approx \expt{p}_0/m$ associated with the 
highest-momentum component, the fractional abundance contribution
$\Omega_0/\Omega$ associated with this same component, and the ratio 
\il{\expt{v}_1/\expt{v}_0 \approx \expt{p}_1/\expt{p}_0 < 1} of 
the average velocities of the two components.  Moreover, we shall take  
\il{\sigma_0 = \sigma_1 = 0.63}, such that these widths accord
with the standard deviation of $\log p$ for our baseline velocity 
distribution in Eq.~\eqref{eq:gthermal}. 

\begin{figure}[t]
\begin{center}
    \includegraphics[width=0.49\textwidth, keepaspectratio]{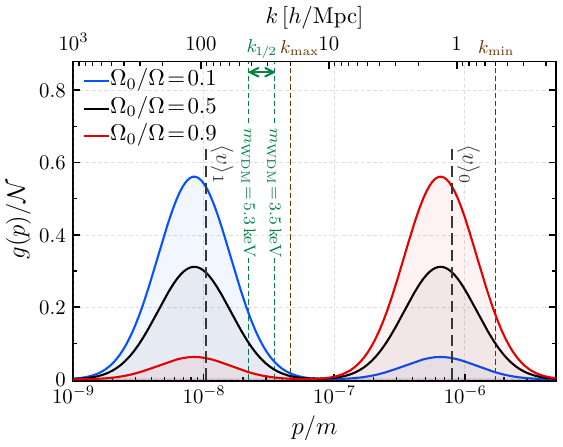}
        \caption{
            Three $g(p)$ distributions of the general form 
            specified in Eq.~\eqref{eq:gsum}, each of which is a 
            superposition of \il{N_G = 2} Gaussian peaks.  
            The average momenta $\expt{p}_0$ and $\expt{p}_1$ associated
            with these two peaks are the same for all three distributions,
            but the fractional abundance $\Omega_0/\Omega$ associated with 
            the higher-momentum peak is different for each one.
            The value of $k$ above which the dark-matter particles in 
            $g(p)$ with velocity $p/m$ can suppress power due 
            to free-streaming effects is indicated along the top axis.
            The dashed brown vertical lines indicate the wavenumbers  
            $\kmin$ and $\kmax$ which delimit the range of $k$ values 
            over which the averaging is performed
            in Eq.~\eqref{eq:dA} in order to obtain $\dA$.
            The dashed green vertical lines indicate the values of 
            $\khalf$ which correspond to the WDM masses 
            \il{\mWDM = 3.5~{\rm keV}} and \il{\mWDM = 5.3~{\rm keV}}.
            }
            \label{fig:gtwinpeaks}
\end{center}
\end{figure}

In Fig.~\ref{fig:gtwinpeaks}, we show three examples of double-peak 
$g(p)$ distributions, all with \il{\expt{v}_0 = 8.0 \times 10^{-7}} 
and \il{\expt{v}_1/\expt{v}_0 = 1.3 \times 10^{-2}}, 
but with different values of $\Omega_0/\Omega$.  We plot 
these distributions as functions of $p/m$.  In order to provide
a sense of how different parts of these $g(p)$ distributions impact 
structure formation, we also associate a wavenumber $k$ with each 
value of $p/m$.  This wavenumber, which is indicated 
on the top axis of the figure, effectively represents the value 
of $k$ above which a population 
of dark-matter particles all moving at exactly the same speed 
$v\approx p/m$ contributes to the suppression of $P(k)$ through 
free-streaming, according to Eq.~\eqref{eq:dFSH}.
The dashed green vertical lines indicate the values of $\khalf$ 
which correspond to the WDM masses \il{\mWDM = 3.5~{\rm keV}} and 
\il{\mWDM = 5.3~{\rm keV}}.  The \Lya\ constraint derived from 
the half-mode recast for either of these $\mWDM$ values  
is sensitive to the detailed shape of $g(p)$ in the region of the plot 
to the right of the corresponding $\khalf$ line.  By contrast,
the dashed brown vertical lines in the figure delimit the range of 
wavenumbers \il{\kmin \leq k \leq \kmax} over which the averaging is 
performed in Eq.~\eqref{eq:dA} in order to obtain $\dA$.  The \Lya\
constraint derived from the $\dA$ recast is sensitive to the 
detailed of $g(p)$ in the region to the left of the $\kmin$ line.

\begin{figure*}[t]
    \begin{center}
    \includegraphics[width=1.0\textwidth, keepaspectratio]{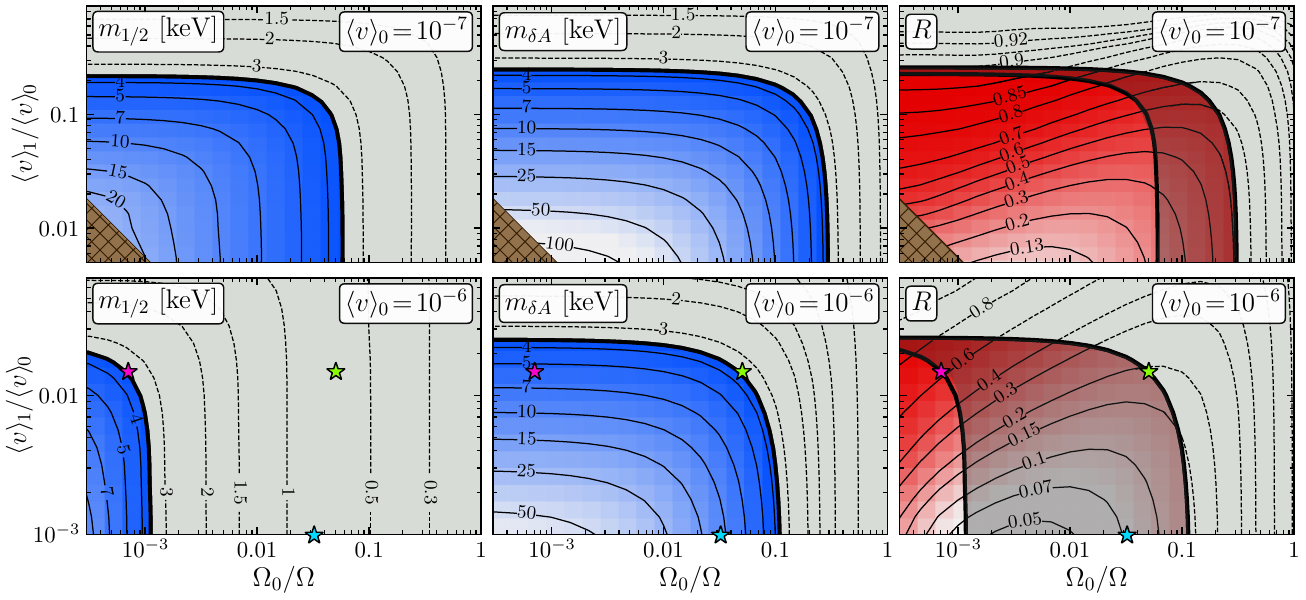}
        \caption{
            Contours in the $(\Omega_0/\Omega,\expt{v}_1/\expt{v}_0)$-plane 
            of $\mhalf$ (left panel of each row) and $\mdA$ (center panel), 
            as well as the comparator $R$ (right panel), for the general 
            phase-space distribution in Eq.~\eqref{eq:gsum} with \il{N_G = 2}.  
            The results shown in the top and bottom rows respectively 
            correspond to the choices \il{\expt{v}_0 = 10^{-7}} and
            \il{\expt{v}_0 = 10^{-6}} for the average velocity of
            the higher-velocity Gaussian peak, while the widths of 
            both peaks have been set to \il{\sigma_0 = \sigma_1 = 0.63} 
            in all panels.  As in Fig.~\protect\ref{fig:unimodal}, the 
            thick black curve in the left and center panels of each row 
            indicates the \Lya\ constraint obtained by employing the
            bound \il{\mWDM \gtrsim 3.5\,{\rm keV}} in implementing 
            the corresponding recast.  The gray region above and to 
            the right of each thick black curve  is excluded.  The 
            exclusion contours for both of these recasts are superimposed 
            on the contours of $R$ shown in the right panel. 
            The magenta and green stars indicate the combinations of 
            $\Omega_0/\Omega$ and $\expt{v}_1/\expt{v}_0$ for which the 
            transfer functions are plotted in 
            Figs.~\protect\ref{fig:twinpeaksPk} and~\protect\ref{fig:compareabg}.  
            The blue star indicates an additional combination of $\Omega_0/\Omega$ 
            and $\expt{v}_1/\expt{v}_0$ for which the transfer function is also
            plotted in Fig.~\protect\ref{fig:compareabg}.
            Within the brown cross-hatched regions of the panels in the upper row, 
            $g(p)$ is consistent with \Lya\ constraints, but the values of 
            $\mhalf$ and $\mdA$ could not both be reliably determined 
            numerically to the desired accuracy.
            }
            \label{fig:twinpeaks}
    \end{center}
\end{figure*}
    
\begin{figure}[t]
    \begin{center}
    \includegraphics[width=0.49\textwidth, keepaspectratio]{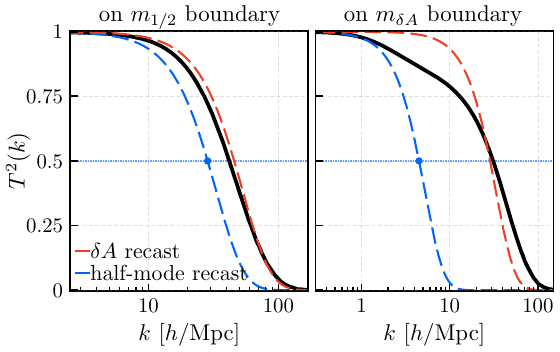}
        \caption{
            The transfer functions $T^2(k)$ which correspond to the 
            parameter-space points indicated by the magenta and green stars in
            Fig.~\ref{fig:twinpeaks}.  The results in the left panel correspond 
            to the magenta star, which is situated on the \Lya\ exclusion contour 
            obtained from the half-mode recast, while the results in the right 
            panel correspond to the green star, which is situated on the \Lya\ 
            exclusion contour obtained from the $\dA$ recast.  The solid black 
            curve in each panel represents the result of a numerical calculation, 
            while the dashed curves represent the WDM transfer functions onto 
            which this $T^2(k)$ is effectively mapped by the half-mode and 
            $\dA$ recasts. 
            }
            \label{fig:twinpeaksPk}
    \end{center}
\end{figure}

Proceeding as we did with the Class-I distributions that we examined 
in Sect.~\ref{sec:minimal}, we survey the parameter space of possible
double-peak $g(p)$ distributions, 
evaluating the critical masses $\mhalf$ and $\mdA$ and the 
comparator $R$ at each point.  The results of this
analysis are shown in Fig.~\ref{fig:twinpeaks}.  In particular,
we show contours of $\mhalf$ (left panel of each row), $\mdA$ 
(center panel), and $R$ (right panel) in the
$(\Omega_0/\Omega,\expt{v}_1/\expt{v}_0)$-plane for two different
choices of $\expt{v}_0$.  The results displayed in the top row 
correspond to the choice \il{\expt{v}_0 = 10^{-7}}, while those 
displayed in the bottom row correspond to the choice 
\il{\expt{v}_0 = 10^{-6}}.  In both rows, 
we have fixed \il{\sigma_0 = \sigma_1 = 0.63}.  
As in Fig.~\protect\ref{fig:unimodal}, the thick black 
curves in the left and center panels of each row of the figure 
indicate the \Lya\ constraints obtained by employing the
bound \il{\mWDM \gtrsim 3.5\,{\rm keV}} in implementing 
the corresponding recast.  The gray region above and to 
the right of each such curve is excluded.  
The exclusion contours for both of these recasts 
are superimposed on the contours of $R$ shown in the right panel
of the corresponding row.  Within the brown cross-hatched regions 
of the figure, $g(p)$ is consistent with \Lya\ constraints, 
but the values of $\mhalf$ and $\mdA$ could not both be reliably 
determined numerically to the desired accuracy.

The shapes of the exclusion contours in Fig.~\ref{fig:twinpeaks} reveal
that \Lya\ data effectively impose two separate constraints on these 
double-peak $g(p)$ distributions.  First, they place an upper bound on the
abundance $\Omega_0$ associated with the higher-velocity 
peak.  Second, they also place an upper bound on the average velocity 
$\expt{v}_1$ associated with the lower-velocity peak.  The precise upper 
limit on each of these quantities depends on the value of $\expt{v}_0$.
It is also worth noting that the entire right edge of each panel in the 
top row of the figure --- an edge along which $\mhalf$ and $\mdA$ are 
both constant --- corresponds to the single point along the dashed yellow 
line in Fig.~\protect\ref{fig:unimodal} at which $\expt{v}_0 = 10^{-7}$. 

More striking, however, is the disagreement between the results derived
from the half-mode and $\dA$ recasts for Class-II velocity distributions
of this sort.  For \il{\vj{0} = 10^{-7}}, the exclusion region obtained
from the half-mode recast extends beyond the region obtained for the $\dA$
recast by roughly a factor of five in the $\Omega_0/\Omega$ direction.
For \il{\vj{0} = 10^{-6}}, the difference can be over two orders of 
magnitude, and nearly the entire region of the 
$(\Omega_0/\Omega,\expt{v}_1/\expt{v}_0)$-plane shown in the figure is 
excluded according to the $\mhalf$ recast.  This disagreement is also 
reflected in the values of the comparator.  We observe that the values of 
$R$ across a significant portion of the parameter space shown in 
Fig.~\ref{fig:twinpeaks} are quite extreme in comparison with any value 
of $R$ which we encountered for the Class-I velocity distributions in 
Sect.~\ref{sec:minimal}.~ We also observe that $R$ is not a monotonic 
function of $\Omega_0/\Omega$, but rather falls, reaches a minimum, and 
then rises again as this parameter is varied along any line of constant 
$\expt{v}_1/\expt{v}_0$. 

The results in Fig.~\ref{fig:gtwinpeaks} actually illustrate one 
way in which these disagreements can arise for Class-II distributions.  
The vast majority of the abundance associated with the high-velocity peak 
in the $g(p)$ distributions shown lies within the range of $p/m$ within 
which both the half-mode and $\dA$ recasts are sensitive.  However, 
the majority of the abundance associated with the low-velocity peak
lies within the region to which the $\dA$ recast is sensitive, but
the half-mode recast is not.  We might therefore anticipate 
disagreements between the recasts to arise in situations in which
$\Omega_0$ and $\Omega_1$ are both sizable.

Another way to understand the origin of the disagreement between 
these two recasts is to consider how the shape of $T^2(k)$ 
varies across our parameter space.  In Fig.~\ref{fig:twinpeaksPk}, we 
show the numerical results for $T^2(k)$ (solid black curves) for two 
different points within that space.  The curve in the left panel 
corresponds to the point $(\Omega_0/\Omega,\expt{v}_1/\expt{v}_0) 
\approx (7.0\times 10^{-4},0.015)$ indicated by the magenta star in 
Fig.~\ref{fig:twinpeaks}, while the curve in the right panel corresponds 
to the point $(\Omega_0/\Omega,\expt{v}_1/\expt{v}_0) 
\approx (4.8\times 10^{-2},0.015)$ indicated by the green star.
In each panel, alongside the true $T^2(k)$ curve we also plot 
the $T^2(k)$ curves obtained for WDM models with $\mWDM$ set 
equal to the corresponding value of $\mhalf$ (dashed blue curve) 
or $\mdA$ (dashed red curve).

The true $T^2(k)$ curve shown in the left panel of 
Fig.~\ref{fig:twinpeaksPk} represents a relatively mild
but nevertheless meaningful departure from the transfer functions
to which the Class-I distributions that we examined in 
Sect.~\ref{sec:minimal} give rise.
The free-streaming of dark-matter particles associated with 
the higher-velocity peak leads to a suppression of power 
at lower $k$.  Since \il{\Omega_0/\Omega \approx 7.0\times 10^{-4}}
is quite small in this case, this suppression is barely discernible 
to the eye.  Nevertheless, it has a non-negligible impact on $\mhalf$,
which is quite sensitive to the behavior of $T^2(k)$ at low $k$.
By contrast, $\mdA$ is far less sensitive to such behavior,
and the corresponding WDM transfer function accords reasonably
well with the true form of $T^2(k)$.  As a result, a 
disagreement arises between the values of $\mhalf$ and $\mdA$.

\begin{figure*}[t]    
    \begin{center}
    \includegraphics[width=0.67\textwidth, keepaspectratio]{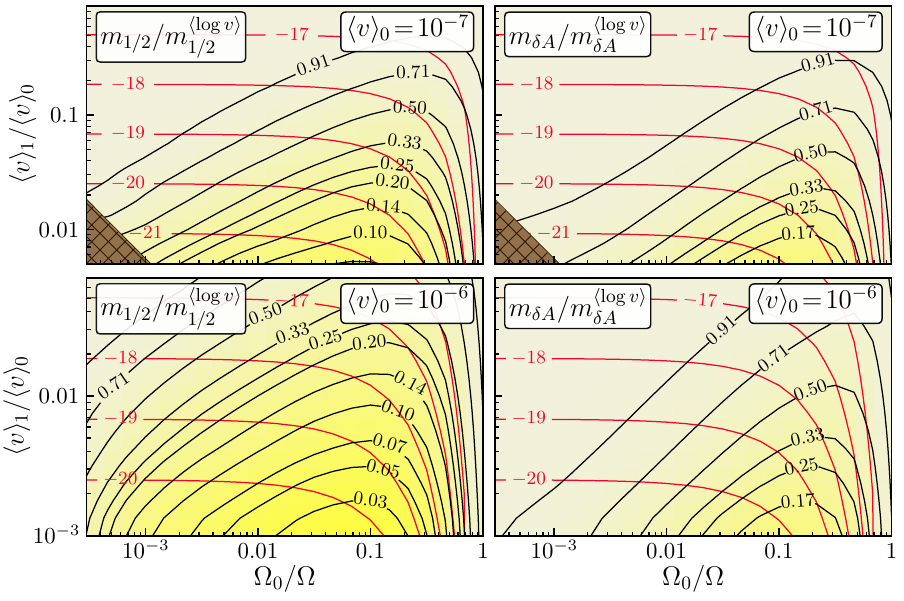}
        \caption{
            The effect of the detailed shape of $g(p)$ on the critical masses 
            $\mhalf$ and $\mdA$.  The left panel of each row shows contours 
            (black curves) in the $(\Omega_0/\Omega,\expt{v}_1/\expt{v}_0)$-plane 
            of the ratio $m_{1/2}/m_{1/2}^{\langle\log v\rangle}$ of the half-mode mass 
            obtained for a $g(p)$ distribution of the form in Eq.~\eqref{eq:gsum} 
            with \il{N_G = 2} and \il{\sigma_0 = \sigma_1 = 0.63} to the half-mode 
            mass obtained for the $g(p)$ distribution with \il{N_G = 1} and 
            \il{\sigma_0 = 0.63} which has the same value of $\langle\log v\rangle$.  
            In the right panel of each row, we show contours of the corresponding ratio 
            $m_{\delta A}/m_{\delta A}^{\langle\log v\rangle}$ for the 
            other critical mass.
            Contours (red curves) of $\langle \log v\rangle$ itself are also included 
            in each panel.  The results shown in the top and bottom rows 
            correspond to the choices \il{\expt{v}_0 = 10^{-7}} and
            \il{\expt{v}_0 = 10^{-6}}, respectively.  We observe that 
            $m_{1/2}/m_{1/2}^{\langle\log v\rangle}$ and 
            $m_{\delta A}/\mdA^{\langle\log v\rangle}$ can each differ significantly from 
            unity, even within the region parameter space which is consistent with the 
            corresponding exclusion contour in Fig.~\protect\ref{fig:twinpeaks}.
            }
            \label{fig:twinpeaksextra}
    \end{center}   
\end{figure*}

The true $T^2(k)$ curve shown in the right panel represents an even more 
dramatic departure from the transfer functions to which Class-I velocity 
distributions give rise.  The abundance 
associated with the higher-velocity peak is significantly larger
in this case, and the free-streaming of dark-matter particles 
associated with this peak has a more pronounced effect on $T^2(k)$ 
at lower $k$.  As a result, the value of $\mhalf$ is comparatively
low.  However, since the lower-velocity peak carries 
the vast majority of the overall dark-matter abundance, 
$T^2(k)$ initially decreases only gradually with increasing $k$.
Indeed, it is only at substantially higher values of $k$ that 
the free-streaming of the dark-matter particles associated with
the low-velocity peak contributes to the suppression of 
structure and induces a precipitous drop in $T^2(k)$.  As a
result, the value of $\mdA$ is comparatively high, and a
significant disagreement arises between the results of our 
two recasts.

Fig.~\ref{fig:twinpeaks}, especially when supplemented by 
Fig.~\ref{fig:twinpeaksPk}, 
reveals a great deal of information about the way in which the results of 
the half-mode and $\dA$ recasts are modified by the presence of an additional 
Gaussian peak in $g(p)$.  However, since the center of 
velocity $\expt{\log v}_{01}$ for $g(p)$ in the \il{N_G = 2} case 
depends on both $\expt{v}_1/\expt{v}_0$ and $\Omega_0/\Omega$, this 
figure does not reveal to what extent the variation of $\mhalf$ and $\dA$ 
across the parameter space shown is simply a consequence of the variation in 
$\expt{\log v}_{01}$ and to what extent it is in fact a consequence 
of variations in the detailed shape of $g(p)$.  As such,
in Fig.~\ref{fig:twinpeaksextra}, we provide additional information which
illustrates how the detailed shape of $g(p)$ affects our results for
these critical masses across the same region of the
$(\Omega_0/\Omega,\expt{v}_1/\expt{v}_0)$-plane
shown in Fig.~\ref{fig:twinpeaks}.  In the left panel of 
Fig.~\ref{fig:twinpeaksextra}, we show contours (black curves) of the ratio 
$m_{1/2}/m_{1/2}^{\langle\log v\rangle}$ of the half-mode mass 
obtained for the resulting $g(p)$ distribution to the half-mode 
mass obtained for the $g(p)$ distribution with \il{N_G = 1} and 
\il{\sigma_0 = 0.63} which has the same center of velocity.  
In the right panel of each row, we show contours of the corresponding 
ratio $m_{\delta A}/m_{\delta A}^{\langle\log v\rangle}$ for the 
other critical mass.  In each panel of the figure, we also  
show contours (red curves) of the center of velocity $\expt{\log v}_{01}$ 
itself.

The ratios $m_{1/2}/m_{1/2}^{\langle\log v\rangle}$ and
$m_{\delta A}/m_{\delta A}^{\langle\log v\rangle}$ characterize the 
impact that the detailed shape of $g(p)$ has on the results of 
the two recasts.  When either of these ratios departs significantly 
from unity, the result obtained from the corresponding recast differs
significantly from the na\"{i}ve result that one would obtain 
from a Class-I distribution with the same center of velocity.
We see from Fig.~\ref{fig:twinpeaksextra} that indeed both ratios
depart significantly from unity across large regions of the
$(\Omega_0/\Omega,\expt{v}_1/\expt{v}_0)$-plane.  The most 
significant departures occur when $\expt{v}_1/\expt{v}_0$ is small
and $\Omega_0/\Omega$ is $\mathcal{O}(0.1)$.  

We also observe that $m_{1/2}/m_{1/2}^{\langle\log v\rangle}$ and
$m_{\delta A}/m_{\delta A}^{\langle\log v\rangle}$ are both less than
unity throughout the entirety of the 
$(\Omega_0/\Omega,\expt{v}_1/\expt{v}_0)$-plane.
This is ultimately a reflection of the fact that the bimodal $g(p)$ distribution 
associated with each point in this plane can be viewed as a bifurcation of the 
unimodal $g(p)$ distribution to which it is being compared --- a bifurcation of
precisely the sort described in the procedure outlines at the beginning of this
section.  This bifurcation shifts some portion of the abundance associated with 
the original peak to higher $v$ and some portion to lower $v$.  However, because   
$\mhalf$ and $\mdA$ are in general more sensitive to modifications 
of the dark-matter velocity distribution at high $v$ than at low $v$, 
this shift leads to a decrease in both critical-mass ratios.
Moreover, as we observe from the figure, the smaller $\expt{v}_1/\expt{v}_0$ is 
for a given $\Omega_0/\Omega$, the more significant that decrease is.
We note that this is due not to any decrease in these critical masses   
themselves --- indeed, $\mhalf$ and $\mdA$ both actually {\it increase}\/ as 
$\expt{v}_1/\expt{v}_0$ decreases, as is evident from Fig.~\ref{fig:twinpeaks} ---
but rather to the fact that the corresponding critical masses 
$m_{1/2}^{\langle\log v\rangle}$ and $m_{\delta A}^{\langle\log v\rangle}$ 
increase more quickly.
We also note that the decrease in $m_{1/2}/m_{1/2}^{\langle\log v\rangle}$ 
for a given set of model parameters is typically more extreme than the 
decrease in $m_{\delta A}/m_{\delta A}^{\langle\log v\rangle}$,
owing to the fact that $m_{1/2}$ is more sensitive to small changes in
the dark-matter velocity distribution at high $v$ than $\mdA$ is.

We remark that in defining the ratios $m_{1/2}/m_{1/2}^{\langle\log v\rangle}$ 
and $m_{\delta A}/m_{\delta A}^{\langle\log v\rangle}$ for a given $g(p)$ 
distribution with \il{N_G = 2}, we have taken as our point of comparison a 
unimodal $g(p)$ distribution which has the same center of 
velocity as the bimodal distribution, but which also has the same width 
$\sigma$ as either of its two individual peaks.  Since we have taken these 
widths, and therefore $\sigma$, to be equal to that of a WDM distribution, 
this comparison allows us to investigate the effect of the detailed shape
of $g(p)$ beyond the first moment --- \ie, the mean --- of this 
distribution.  That said, 
it would also be interesting to compare the results for $\mhalf$ 
and $\mdA$ that we obtain for $N_G=2$ to the corresponding values obtained 
for a unimodal $g(p)$ distribution which not only has the same center of velocity, 
but also has a width equal to the standard deviation of the {\it entire}\/ 
bimodal $g(p)$ distribution in $(\log p)$-space.  Indeed, such a comparison 
would highlight the importance of even higher moments of the $g(p)$ distribution
beyond the second.  However, performing such a comparison is challenging because
interesting regions of the \il{N_G = 2} parameter space correspond to extremely 
large values of $\sigma$ --- values for which the numerical tools we have
used in computing $\mhalf$ and $\mdA$ become insufficient.  We leave such an 
analysis to future work.

\begin{figure*}[t]
    \begin{center}
    \includegraphics[width=1.0\textwidth, keepaspectratio]{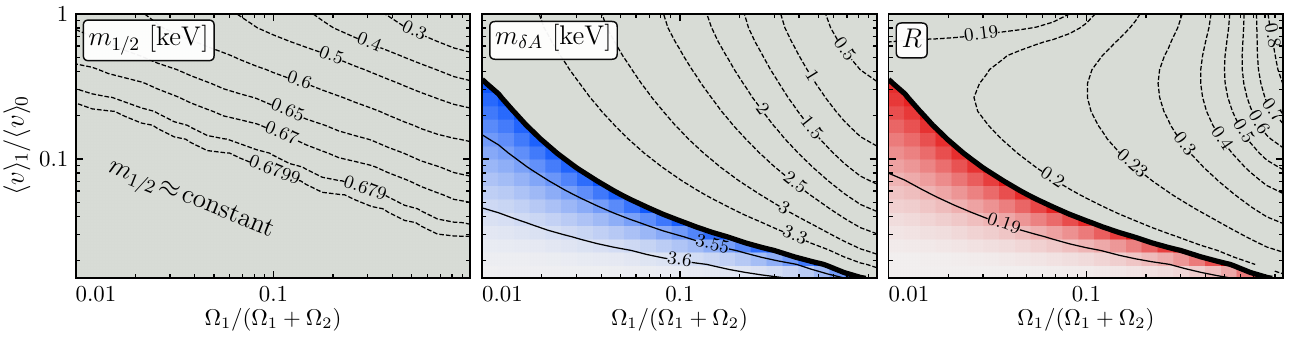}
        \caption{
            Contours in the 
            $(\Omega_1/(\Omega_1+\Omega_2),\expt{v}_1/\expt{v}_0)$-plane 
            of $\mhalf$ (left panel) and $\mdA$ (center panel), as well as 
            the comparator $R$ (right panel), for the general phase-space 
            distribution in Eq.~\eqref{eq:gsum} with \il{N_G = 3}.  The 
            results shown correspond to the choices \il{\expt{v}_0 = 10^{-6}}, 
            \il{\sigma_0 = \sigma_1 = \sigma_2 = 0.63},
            \il{\Omega_0/\Omega = 4.8\times 10^{-2}}, and
            \il{\expt{\log v}_{12} = -18.21}.
            The thick black curves in the center and right panels 
            indicate the \Lya\ constraints obtained by employing the
            bound \il{\mWDM \gtrsim 3.5\,{\rm keV}} in implementing 
            the $\dA$ recast.  The gray regions above and to 
            the right of these constraint contours are excluded.
            }
            \label{fig:threepeaks}
    \end{center}
\end{figure*}

\subsection{The Effects of Additional Modes\label{subsec:threepeaks}}

Our analysis of the bi-modal distributions in the previous section 
constitutes a necessary first step in establishing \Lya\ constraints on 
Class-II velocity distributions.  However, the results of this analysis 
prompt several additional questions.  Perhaps the most important such 
question concerns the manner in which the \Lya\ constraints that we have 
derived for the $g(p)$ distribution in Eq.~\eqref{eq:gsum} with 
\il{N_G = 2} generalize to the case in which \il{N_G > 2}.

Given that a systematic survey of the full parameter space of possible 
$g(p)$ distributions is impractical for \il{N_G > 2}, we shall 
investigate the answer to this question 
according to the procedure outlined at the beginning of this 
section.  We begin by identifying an illustrative benchmark point within
the \il{N_G = 2} parameter space that we examined above.  We then
examine the subspace of the full \il{N_G=3} parameter space associated 
with bifurcations of the lowest-velocity peak in the $g(p)$ 
distribution for this benchmark for which the center of velocity 
$\expt{\log v}_{12}$ is the same as it was for that original peak.
We adopt as our parameter-space benchmark a point 
which lies directly on one of the exclusion contours in 
Fig.~\ref{fig:twinpeaks}.  In particular, we choose the point 
\il{(\Omega_0/\Omega, \vj{1}/\vj{0}) = (4.8\times 10^{-2},0.015)} 
indicated by the green star appearing in the bottom panels of 
Fig.~\ref{fig:twinpeaks}.  We take the two free parameters which
describe the possible bifurcations of the low-velocity peak in
this distribution consistent with our condition on $\expt{\log v}_{12}$  
to be $\expt{v}_1/\expt{v}_0$ and $\Omega_1/(\Omega_1 + \Omega_2)$.

In Fig.~\ref{fig:threepeaks}, we show contours of $\mhalf$ (left panel),
$\mdA$ (center panel), and $R$ (right panel) in the
$(\Omega_1/(\Omega_1+\Omega_2),\expt{v}_1/\expt{v}_0)$-plane for 
the values of $\Omega_0/\Omega$, $\expt{v}_0$, $\sigma_0$, $\sigma_1$, 
$\sigma_2$, and $\expt{\log v}_{12}$ specified above.  
The bottom right corner of each panel corresponds to the location 
of our benchmark point within this space. 
The thick black curves in the center and right panels represent 
the \Lya\ exclusion contours obtained by employing the bound 
\il{\mWDM \gtrsim 3.5\,{\rm keV}} in implementing the $\dA$ recast.
The gray regions above and to the right of these contours are 
excluded.  We observe that each of these contours terminates on and 
includes our benchmark.  Indeed, this is to be expected, given that this 
benchmark corresponds to a point which lies along the 
exclusion contour in Fig.~\ref{fig:twinpeaks} obtained from the 
same recast.  By contrast, we find that the entirety of the parameter 
space shown in Fig.~\ref{fig:threepeaks} is excluded according to the 
half-mode recast.  This is not unexpected either, given that our 
benchmark lies well within the region of the
$(\Omega_0/\Omega,\expt{v}_1/\expt{v}_0)$-plane in 
Fig.~\ref{fig:twinpeaks} which is excluded according to this same 
recast.

Generally speaking, we observe from Fig.~\ref{fig:threepeaks}
that increasing either the relative abundance 
$\Omega_1/(\Omega_1+\Omega_2)$ or the average velocity 
$\vj{1}$ of the intermediate-velocity peak decreases both $\mhalf$ 
and $\mdA$ and thus increases the tension with \Lya\ data.  
These results accord well with our qualitative expectations, 
given that we are holding $\Omega_1 + \Omega_2$ and $\expt{\log v}_{12}$
fixed as we vary these other quantities.  Indeed, increasing either
$\Omega_1/(\Omega_1+\Omega_2)$ or $\vj{1}$ increases the
differential number density of dark-matter particles at higher $v$
and enhances the suppression of power at lower $k$ due to 
free-streaming effects.  Moreover, the shapes of the contours in the left 
two panels of Fig.~\ref{fig:threepeaks} provide interesting additional 
information about how the results of the two recasts depend on
the properties of the two lower-velocity peaks.  In particular, 
they indicate that the values of both $\mhalf$ and $\mdA$ are more 
sensitive to changes in $\expt{v}_1/\expt{v}_0$ than they are to 
$\Omega_1/(\Omega_1+\Omega_2)$, at least in the regime in which latter 
quantity is sizable.  Indeed, it is only for 
\il{\Omega_1/(\Omega_1+\Omega_2) \lesssim 0.05} that the slope of the 
exclusion contour associated with the $\dA$ recast steepens.

Most importantly, however, we see from Fig.~\ref{fig:threepeaks} that 
our comparator $R$ can again differ significantly from unity, and moreover 
does so precisely within the region of parameter space which is considered 
viable according to our recasts.   This reinforces the cautionary 
note --- already observed from Fig.~\ref{fig:twinpeaks} --- that our 
recasts can yield \Lya\ constraints which differ significantly from 
each other for complex, multi-modal dark-matter phase-space distributions.

\section{Comparison with Parametric-Function Techniques\label{sec:abg}}


Our primary focus in this paper has been to illustrate the pitfalls which can 
arise when the half-mode and $\delta A$ recasts are employed in order to 
estimate \Lya\ constraints on dark-matter scenarios with Class-II velocity 
distributions.
Of course, there exists an alternative method for estimating \Lya\ 
constraints on non-cold dark-matter scenarios~\cite{Murgia:2017lwo}.  
This method involves positing a simple, parametric model for $T^2(k)$ 
and performing numerical simulations in order to evaluate the 
corresponding flux power spectra at a representative sample of points 
within the parameter space of that model.  These spectra can then be 
compared to observed flux power spectra in order to determine which 
regions of that parameter space are consistent with \Lya\ data and 
which are excluded.  Once a survey of this sort has been performed,  
constraints on a given dark-matter scenario can be estimated by fitting 
the transfer function obtained for that scenario to the parametric model.  
If the best-fit values for the parameters of that model lie within an
excluded region, the dark-matter scenario is excluded.  

Such a survey was performed in Ref.~\cite{Murgia:2018now} using 
the parametric function~\cite{Murgia:2017lwo}
\begin{equation}
  T^2(k) ~=~ [1 + (\alpha k)^\beta ]^{2\gamma}~,
  \label{eq:alphabetagamma}
\end{equation}
where $\alpha \geq 0$, $\beta > 0$, and $\gamma > 0$ are the free 
parameters which characterize the model.  The authors of
Ref.~\cite{Murgia:2018now} employed hydrodynamic simulations in order to 
place \Lya\ constraints on the three-dimensional parameter space of this
model based on a scan over a large number of combinations of 
$\alpha$, $\beta$, and $\gamma$.

While the parametric function in Eq.~(\ref{eq:alphabetagamma}) is 
capable of modeling a broad range of transfer functions quite accurately, in general
it is not capable of modeling accurately the kinds of transfer functions 
that arise from Class~II velocity distributions.  Indeed, as was shown in 
Eqs.~(3.19) -- (3.21) of Ref.~\cite{Dienes:2020bmn}, 
the matter power spectra described by this parametric function are 
characteristic of strictly unimodal $g(p)$ 
distributions.\footnote{After this paper was originally submitted for 
publication, Ref.~\cite{Hooper:2022byl} appeared in which this 
parametrization of the transfer function was extended
to include an additional parameter $\delta$.  While this extension 
extends the applicability of this parametrization to a wider variety
of situations (including the possibility of an additional, purely 
cold dark-matter component), the resulting $T^2(k)$ function continues
to correspond to that of a single peak, as can be verified using the 
same techniques as were employed in analyzing the 
$\{\alpha,\beta,\gamma\}$ parametrization in Eqs.~(3.19) -- (3.21) of 
Ref.~\cite{Dienes:2020bmn}.  The existence of this additional 
parametrization therefore does not change any of our conclusions.}

\begin{figure*}[t!]
    \begin{center}
    \includegraphics[width=0.7\textwidth, keepaspectratio]{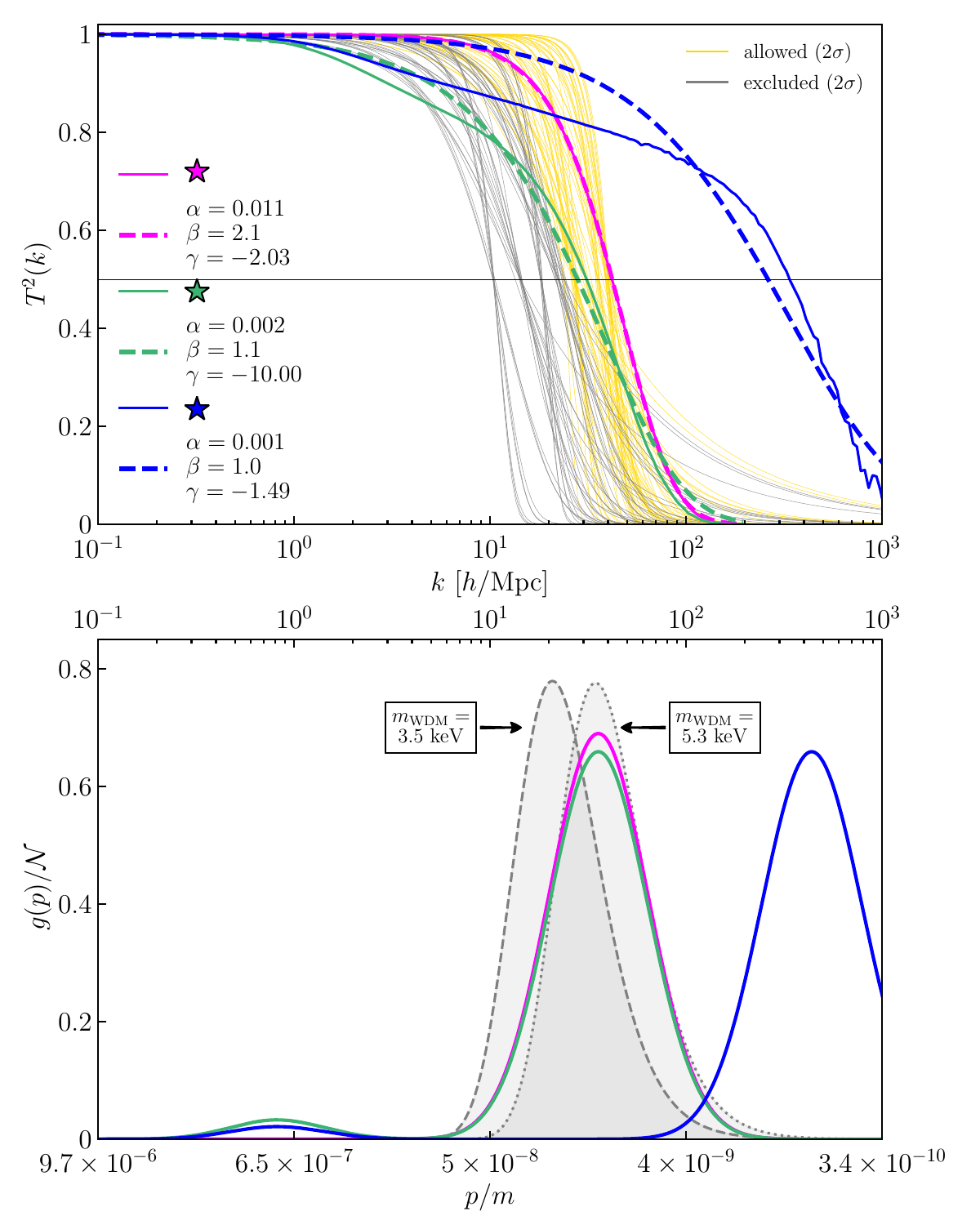}
        \caption{
          {\it Top panel}\/: The solid magenta and green curves represent the transfer 
          functions obtained for the parameter-space points indicated by the 
          stars of the corresponding colors in Fig.~\protect\ref{fig:twinpeaks},
          while the solid blue curve represents the transfer function obtained 
          for the parameter-space point with the same $\expt{v}_0$, $\sigma_0$,
          and $\sigma_1$ values as these other two points, but with 
          $\Omega_0/\Omega = 0.032$ and $\expt{v}_1/\expt{v}_0 = 10^{-3}$.
          The dashed curves of the corresponding colors represent the best-fit
          functions of the form given in Eq.~(\ref{eq:alphabetagamma}) to these
          three curves.  The best-fit values of the parameters $\alpha$, $\beta$,
          and $\gamma$ in each case are provided in the legend.  The yellow and
          gray curves represent the transfer functions of the form given in 
          Eq.~(\ref{eq:alphabetagamma}) for the combinations of $\alpha$, $\beta$, 
          and $\gamma$ included in the survey performed in Ref.~\cite{Murgia:2018now}.
          The yellow curves represent the transfer functions which were found to be 
          consistent with \Lya constraints, while the gray curves represent the 
          transfer functions which were found to be excluded.
          {\it Bottom panel}\/: The $g(p)$ distributions which give rise to the $T^2(k)$ 
          curves of the corresponding colors in the top panel. 
          The value of $k$ above which the dark-matter particles in $g(p)$ with 
          velocity $p/m$ can suppress power due to free-streaming effects is 
          indicated by the tick marks on the top axis.  These $g(p)$ distributions
          are displayed such that these tick marks are uniformly spaced 
          on a logarithmic scale and such that $k$ increases from left to right.
          The corresponding $p/m$ values indicated on the bottom axis are
          therefore spaced non-uniformly and increase from right to left.  The
          $g(p)$ distributions for WDM models with $\mWDM = 3.5$~keV and 
          $\mWDM = 5.3$~keV, which are indicated by the dashed and dotted gray 
          curves, respectively, are also included for reference. 
            \label{fig:compareabg}}
    \end{center}
\end{figure*}

The implications that this property of the parametric function in
Eq.~(\ref{eq:alphabetagamma}) has for establishing \Lya\ constraints on dark-matter
models with Class~II velocity distributions are perhaps best conveyed by means of 
concrete examples.  In the top panel of Fig.~\ref{fig:compareabg}, we show the transfer 
functions (solid magenta, green, and blue curves) obtained for the three parameter-space 
points indicated by the stars of the corresponding colors in the bottom panels of 
Fig.~\ref{fig:twinpeaks}.  The the point indicated by the blue star 
corresponds to the parameter assignments 
$(\Omega_0/\Omega,\expt{v}_1/\expt{v}_0) = (0.032,10^{-3})$.  
This point, which lies within the allowed region in Fig.~\ref{fig:twinpeaks} associated 
with the $\delta A$ recast, is excluded according to the $m_{1/2}$ recast. 
However, the corresponding transfer function represents a far more dramatic departure 
from the transfer functions characteristic of Class-I velocity distributions than 
do the transfer functions obtained for the other two parameter-space points.  
In the bottom panel, we show the $g(p)$ distributions for these three 
parameter-space points.  We note that the ranges of $k$ associated with the individual 
peaks in each of these $g(p)$ distributions align with the ranges of $k$ within which 
the corresponding $T^2(k)$ curve exhibits a significant decrease in its slope.  
The relationship between features in $g(p)$ and $T^2(k)$ suggested 
by this juxtaposition is articulated more concretely in Ref.~\cite{Dienes:2020bmn}.

In the top panel of Fig.~\ref{fig:compareabg}, we also show the transfer function 
(dashed curve of the corresponding color) which represents the 
best fit to the true $T^2(k)$ curve for each of three parameter-space points 
using the parametrization in Eq.~(\ref{eq:alphabetagamma}).  
The values of $\alpha$, $\beta$, and $\gamma$
for each of these best-fit curves are provided in the legend.
The yellow and gray curves represent the transfer functions of the form given in 
Eq.~(\ref{eq:alphabetagamma}) for the combinations of $\alpha$, $\beta$, 
and $\gamma$ included in the parameter-space survey performed in 
Ref.~\cite{Murgia:2018now}.  The yellow curves represent the transfer functions 
which were found to be consistent with \Lya\ constraints, while the gray 
curves represent the transfer functions which were found to be excluded. 

For the magenta $g(p)$ distribution --- a distribution for which the relative 
abundance $\Omega_0/\Omega \approx 7.0 \times 10^{-4}$ of the higher-velocity 
peak is quite small --- we observe that the parametrization in 
Eq.~(\ref{eq:alphabetagamma}) provides an excellent fit to the corresponding 
transfer function.  Since this transfer function lies well within the region 
spanned by the yellow curves, it is clear that this distribution would not be 
excluded by \Lya\ constraints according to the simulations performed in 
Ref.~\cite{Murgia:2018now}.  According to the the half-mode recast, this  
$g(p)$ distribution would be marginal, while according to the $\delta A$
recasts, it would be allowed, as shown in Fig.~\ref{fig:twinpeaks}.  In
this case, then, the results obtained using the parametric function in
Eq.~(\ref{eq:alphabetagamma}) are more consistent with the results of 
the $\delta A$ recast. 

By contrast, for the green $g(p)$ distribution, the best fit to the 
corresponding transfer function using the functional parametrization in 
Eq.~(\ref{eq:alphabetagamma}) differs more significantly from the true
$T^2(k)$ curve.  Indeed, for wavenumbers within the range 
$1h/{\rm Mpc}\lesssim k\lesssim20h/{\rm Mpc}$, the relative error can  
be as high as $5\%$.  Nevertheless, it is still reasonable to assume that 
this $g(p)$ distribution is excluded.  Indeed, the $T^2(k)$ curve 
obtained from the fit reflects a significant suppression of power 
within the relevant range of $k$ and lies well within the region spanned 
by the gray curves, while the true $T^2(k)$ curve is even further 
suppressed.  This $g(p)$ distribution would likewise be excluded
by the half-mode recast, while according to the $\delta A$ recast 
it would be marginal.  In this case, then, the results obtained using 
the parametric function in Eq.~(\ref{eq:alphabetagamma}) are more 
consistent with the results of the half-mode recast.

Finally, for the blue $g(p)$ distribution, the best fit to the 
corresponding transfer function using the functional parametrization 
in Eq.~\ref{eq:alphabetagamma} overestimates the true value of 
$T^2(k)$ by as much $12\%$ for wavenumbers within the range 
$1h/{\rm Mpc}\lesssim k\lesssim100h/{\rm Mpc}$.  Based solely on 
the $T^2(k)$ curve obtained from the fit, one would conclude 
that this $g(p)$ distribution is allowed, given that it lies well 
within the region spanned by the yellow curves.  However, the 
true $T^2(k)$ curve lies significantly below this best-fit curve 
throughout most of the relevant range of $k$.  Thus, we cannot 
conclusively determine whether the corresponding $g(p)$ distribution 
is allowed or excluded by comparing with the simulation
results provided in Ref.~\cite{Murgia:2018now}.  Thus, for Class~II 
velocity distributions of this sort, Lyman-$\alpha$ constraints 
obtained from this parametric-function technique are not
necessarily any more reliable than those obtained from the half-mode 
or $\delta A$ recasts.

Taken together, the results displayed in Fig.~\ref{fig:compareabg}
indicate that while the parametric function in Eq.~(\ref{eq:alphabetagamma}) 
may capture the gross features of the transfer functions 
associated with certain Class-II dark-matter velocity distributions (particularly those that resemble Class-I distributions), they do not generically provide an accurate fit to these $T^2(k)$. 
Indeed, the more significantly $g(p)$ departs from the Class-I velocity 
distributions that we examined in Sect.~\ref{sec:minimal}, the less 
accurate the fit to the corresponding $T^2(k)$ becomes.  Thus, for
velocity distributions of this sort, the Lyman-$\alpha$ constraints 
obtained from this parametric function are not necessarily any more 
reliable than those obtained from the half-mode or $\delta A$ 
recasts.  

That said, the results displayed in Fig.~\ref{fig:compareabg} also 
highlight the importance of performing hydrodynamic simulations 
on dark-matter models beyond the baseline WDM model in deriving 
\Lya\ constraints.  The parametric-function technique developed in 
Ref.~\cite{Murgia:2018now} provides a benchmark for establishing such 
constraints on dark-matter models with more general velocity 
distributions.  Indeed, while the parametric function in 
Eq.~(\ref{eq:alphabetagamma}) is not capable of modeling accurately 
the kinds of transfer functions that arise from Class~II velocity 
distributions, an analysis similar to the one performed in 
Ref.~\cite{Murgia:2018now}, but with a more adaptable parametric 
function could in principle extend the applicability of this 
technique to velocity distributions within this class.

\section{Conclusions\label{sec:Conclusions}}


In this paper, we have conducted a preliminary investigation into
the constraints imposed by \Lya-forest data on general 
dark-matter velocity distributions.  In particular, we have investigated 
the bounds obtained from two common methods of recasting the
\Lya\ constraints onto a baseline WDM model --- methods which 
are frequently employed in order to estimate these constraints in 
situations in which it would be impractical to perform the hydrodynamic 
simulations that would otherwise be necessary.
In order to conduct this study, we also introduced two new quantities --- 
our critical mass parameters $\mhalf$ and $\mdA$ --- which enabled us 
to compare the results of these two recasts in a systematic, quantitative way.

After evaluating the \Lya\ constraints on a baseline WDM velocity 
distribution, we applied our methods in order to study 
departures from this baseline distribution.  We began by examining
how \Lya\ constraints are affected by comparatively mild 
departures, such as replacing this distribution with a Gaussian
distribution and varying the width of the distribution.  Even for such 
simple, Class-I velocity distributions, we demonstrated that 
disagreements arose between the results obtained from the half-mode
and $\dA$ recasts.  We then extended this study to encompass dark-matter 
velocity distributions which represent a even more dramatic departures from 
our baseline distribution, including multi-modal distributions consisting
of superpositions of Gaussian peaks.  The impact of free-streaming
on $T^2(k)$ in these cases is more subtle, given that different parts of 
these velocity distributions have different thresholds in $k$ above
which they contribute to the suppression of power.
Even for the simplest possible such distributions, which 
comprise only two Gaussian peaks, we have demonstrated that dramatic
disagreements between $\mhalf$ and $\mdA$ --- and therefore between
the \Lya\ constraints obtained for the two recasts --- can arise. 
We also examined how these results are modified by the 
incorporation of additional peaks into these dark-matter velocity 
distributions.

While the results in Sect.~\ref{sec:nonminimal}
serve to highlight how disagreements between the half-mode 
and $\dA$ recasts arise for Class-II velocity distributions,
we can make no statement as to which of these recasting methods 
provides a more accurate representation of the \Lya\ constraint 
for a given dark-matter velocity distribution without performing 
the necessary hydrodynamic simulations. 
Nevertheless, our findings provide important insights into
how highly non-trivial velocity distributions of this sort are 
constrained by \Lya\ data.  First, our results provide qualitative
information about how the \Lya\ constraints on particular kinds of 
dark-matter velocity distributions are likely to be affected when 
certain features of those distributions are modified.  Second, however, 
our results provide {\it general indications}\/ as to when each of 
these recasts is likely to be reliable and when it is likely to be 
unreliable across the landscape of possible Class-II velocity 
distributions.  In this way, our results serve as a cautionary tale 
concerning the implementation of these recasts and highlight the 
kinds of distributions for which the need for dedicated numerical 
analysis is the most pressing.

Several additional comments are in order.  First, while we have taken a 
step toward exploring the \Lya\ constraints on the landscape of possible 
Class-II dark-matter velocity distributions, this landscape is vast and 
there are a number of areas that merit further exploration.  Indeed, while 
a wide variety of such distributions can be decomposed according to
Eq.~\eqref{eq:gsum} with reasonably fidelity, such a Gaussian 
decomposition cannot capture every feature of a generic function perfectly.  
It does not, for example, permit us to study the extent to which the 
exclusion contours are sensitive to the detailed shapes of the individual 
peaks.

Second, along similar lines, there are other aspects of Class-I velocity 
distributions which merit further exploration as well.
In particular, we would like to know how general modifications of  
the shape of the peak can affect the results of the half-mode and 
$\dA$ recasts.  This shape may be quantitatively characterized
by its different moments --- the average momentum, standard deviation, 
skewness, kurtosis, \etc~  We have examined the effect of varying 
the first two of these moments in Sect.~\ref{sec:minimal}, but there 
are many reasons to expect that the higher moments also have a 
significant impact on the results of these recasts, particularly for 
a distribution with a sizable width.  
For example, a positively skewed distribution contains a higher 
proportion of low-velocity particles than a symmetric distribution
does.  As a result, the suppression of structure due to 
free-streaming may become particularly pronounced only at smaller 
scales (corresponding to larger $k$).
By contrast, a negatively skewed distribution contains a higher
proportion of high-velocity particles than does a symmetric distribution
does.  As a result, the suppression of structure can become 
significant at even larger distance scales (corresponding to smaller $k$).
Such effects on $T^2(k)$ are likely to have have an impact on the results
of our recasts.  Indeed, as discussed in Ref.~\cite{Dienes:2020bmn}, the 
skewness of the dark-matter velocity distribution is directly connected 
to fundamental properties of the dark-matter production mechanism.

Finally, our study in this paper was performed under the assumption that 
the shape of the dark-matter phase-space distribution is essentially 
fixed well before the time of matter-radiation equality.  For this reason, 
our analysis may not be applicable to scenarios in which $g(p,t)$ is still 
dynamically evolving at subsequent times, when matter perturbations are 
growing at a significant rate.  Such complications may arise for models 
in which the dark matter has non-negligible 
self-interactions~\cite{Huo:2017vef}.  Such complications may also arise 
in multi-component dark-matter models wherein the heavier
dark-matter species are unstable but long-lived and decay 
primarily to final states comprising lighter dark-matter species.
Indeed, this is what occurs within the context of the Dynamical 
Dark Matter framework~\cite{Dienes:2011ja,Dienes:2011sa,Dienes:2012jb}. 
The resulting gradual conversion of mass energy to kinetic energy
within the dark sector can then ``heat up'' the dark matter at late 
times, resulting in a non-trivial modification of the 
dark-matter velocity distribution.  
Indeed, for certain scenarios of this sort, numerical studies have 
been performed in order to investigate the impact of this effect on 
the \Lya\ forest~\cite{Wang:2013rha} and on other aspects of 
small-scale structure~\cite{Wang:2014ina,Cheng:2015dga}.


\begin{acknowledgments}
The research activities of KRD are supported in part by the Department 
of Energy under Grant DE-FG02-13ER41976 / DE-SC0009913 and by the 
National Science Foundation through its employee IR/D program.
The research activities of FH are supported by the International 
Postdoctoral Exchange Fellowship Program, the National Natural Science 
Foundation of China under Grants 12025507, 11690022, 11947302, 12022514, 
11875003 and is also supported by the Strategic Priority Research 
Program and Key Research Program of Frontier Science of the Chinese 
Academy of Sciences under Grants XDB21010200, XDB23010000, ZDBS-LY-7003, 
and the CAS project for Young Scientists in Basic Research YSBR-006.  
The research activities of JK are supported in part by the Science 
and Technology Research Council (STFC) under the Consolidated 
Grant ST/T00102X/1.  The research activities of BT are supported in 
part by the National Science Foundation under Grant PHY-2014104.  The
research activities of HBY are supported in part by the Department of 
Energy under Grant DE-SC0008541 and the John Templeton Foundation under 
Grant 61884.  BT and HBY would also like to acknowledge the hospitality of 
the Kavli Institute for Theoretical Physics (KITP), which is supported in 
part by the National Science Foundation under Grant PHY-1748958.
The opinions and conclusions expressed herein are 
those of the authors, and do not represent any funding agencies.
\end{acknowledgments}




\bibliographystyle{apsrev4-2}
\bibliography{references}

\end{document}